\documentclass[twocolumn]{aastex62}

\usepackage{amsmath}
\usepackage{multirow}
\usepackage{color}
\usepackage{graphicx}

\newcommand{\nii}{[\ion{N}{2}]}
\newcommand{\ha}{H$\alpha$}
\newcommand{\oiii}{[\ion{O}{3}]}
\newcommand{\feii}{\ion{Fe}{2}}
\newcommand{\cii}{[\ion{C}{2}]}
\newcommand{\hi}{\ion{H}{1}}
\newcommand{\naid}{\ion{Na}{1}}

\shorttitle{Molecular and Ionized Outflow in a Main-Sequence, Massive Galaxy at $z=2$}
\shortauthors{Herrera-Camus et al.}

\begin{document}

\title{The Molecular and Ionized Gas Phases of an AGN-driven Outflow in a Typical Massive Galaxy at $z\approx2$}

\correspondingauthor{Rodrigo Herrera-Camus}
\email{rhc@mpe.mpg.de}

\author{R. Herrera-Camus}
\affil{Max-Planck-Institut f\"{u}r Extraterrestrische Physik (MPE), Giessenbachstr., D-85748 Garching, Germany}
\author{L. Tacconi} 
\affil{Max-Planck-Institut f\"{u}r Extraterrestrische Physik (MPE), Giessenbachstr., D-85748 Garching, Germany}
\author{R. Genzel}
\affil{Max-Planck-Institut f\"{u}r Extraterrestrische Physik (MPE), Giessenbachstr., D-85748 Garching, Germany}
\author{N. F{\"o}rster Schreiber}
\affil{Max-Planck-Institut f\"{u}r Extraterrestrische Physik (MPE), Giessenbachstr., D-85748 Garching, Germany}
\author{D. Lutz}
\affil{Max-Planck-Institut f\"{u}r Extraterrestrische Physik (MPE), Giessenbachstr., D-85748 Garching, Germany}
\author{A. Bolatto}
\affil{Department of Astronomy, University of Maryland, College Park, MD 20742-2421, USA}
\author{S. Wuyts}
\affil{Department of Physics, University of Bath, Claverton Down, Bath, BA2 7AY, United Kingdom}
\author{A. Renzini}
\affil{INAF-Osservatorio Astronomico di Padova, Vicolo dell'Osservatorio 5, I-35122 Padova, Italy}
\author{S. Lilly}
\affil{Department of Physics, Institute for Astronomy, ETH Zurich, CH-8093 Zurich, Switzerland}
\author{S. Belli}
\affil{Max-Planck-Institut f\"{u}r Extraterrestrische Physik (MPE), Giessenbachstr., D-85748 Garching, Germany}
\author{H. \"{U}bler}
\affil{Max-Planck-Institut f\"{u}r Extraterrestrische Physik (MPE), Giessenbachstr., D-85748 Garching, Germany}
\author{T. Shimizu}
\affil{Max-Planck-Institut f\"{u}r Extraterrestrische Physik (MPE), Giessenbachstr., D-85748 Garching, Germany}
\author{R. Davies}
\affil{Max-Planck-Institut f\"{u}r Extraterrestrische Physik (MPE), Giessenbachstr., D-85748 Garching, Germany}
\author{E. Sturm}
\affil{Max-Planck-Institut f\"{u}r Extraterrestrische Physik (MPE), Giessenbachstr., D-85748 Garching, Germany}
\author{F. Combes}
\affil{Observatoire de Paris, LERMA, College de France, CNRS, PSL Univ., Sorbonne Univ. UPMC, F-75014, Paris, France}
\author{J. Freundlich}
\affil{Centre for Astrophysics and Planetary Science, Racah Institute of Physics, The Hebrew University, Jerusalem 91904, Israel}
\author{S. Garc\'ia-Burillo}
\affil{Observatorio Astron\'omico Nacional-OAN, Observatorio de Madrid, Alfonso XII, 3, E-28014-Madrid, Spain}
\author{P. Cox}
\affil{CNRS, UMR 7095, Institut d'Astrophysique de Paris, 75014 Paris, France}
\author{A. Burkert}
\affil{Universitt\"{a}s-Sternwarte Ludwig-Maximilians-Universit\"{a}t (USM), Scheinerstr. 1, M\"{u}nchen, D-81679,Germany}
\affil{Max-Planck-Institut f\"{u}r Extraterrestrische Physik (MPE), Giessenbachstr., D-85748 Garching, Germany}
\author{T. Naab}
\affil{Max-Planck Institute for Astrophysics, Karl Schwarzschildstrasse 1, D-85748 Garching, Germany}
\author{L. Colina}
\affil{Centro de Astrobiolog\'ia (CSIC/INTA), Ctra de Torrej\'on a Ajalvir, km 4, 28850, Torrej\'on de Ardoz, Madrid, Spain}
\affil{Cosmic Dawn Center (DAWN), Niels Bohr Institute, University of Copenhagen / DTU-Space, Technical University of Denmark}
\author{A. Saintonge}
\affil{Department of Physics \& Astronomy, University College London, Gower Place, London WC1E 6BT, UK}
\author{M. Cooper}
\affil{Department of Physics and Astronomy, Frederick Reines Hall, University of California, Irvine, CA 92697, USA}
\author{C. Feruglio}
\affil{INAF Osservatorio Astronomico di Trieste, Via G.B. Tiepolo 11, I-34143 Trieste, Italy}
\author{A. Weiss}
\affil{Max-Planck-Institut f\"{u}r Radioastronomie (MPIfR), Auf dem H\"{u}gel 69, D-53121 Bonn, FRG, Germany}

\begin{abstract}
Nuclear outflows driven by accreting massive black holes are one of the main feedback mechanisms invoked at high-$z$ to reproduce the distinct separation between star-forming, disk galaxies and quiescent spheroidal systems. Yet, our knowledge of feedback at high-$z$ remains limited by the lack of observations of the multiple gas phases in galaxy outflows. In this work we use new deep, high-spatial resolution ALMA CO(3-2) and archival VLT/SINFONI \ha\ observations to study the molecular and ionized components of the  AGN-driven outflow in zC400528 ---a massive, main sequence galaxy at $z=2.3$ in the process of quenching. We detect a powerful molecular outflow that shows a positive velocity gradient and extends for at least $\sim$10~kpc from the nuclear region, about three times the projected size of the ionized wind. The molecular gas in the outflow does not reach velocities high enough to escape the galaxy and is therefore expected to be reaccreted. Keeping in mind the various assumptions involved in the analysis, we find that the mass and energetics of the outflow are dominated by the molecular phase. The AGN-driven outflow in zC400528 is powerful enough to deplete the molecular gas reservoir on a timescale at least twice shorter than that needed to exhaust it by star formation. This suggests that the nuclear outflow is one of the main quenching engines at work in the observed suppression of the central star-formation activity in zC400528.
\end{abstract}


\section{Introduction} \label{sec:intro}

Powerful galaxy outflows induced by active galactic nuclei (AGN) have been invoked as one of the main drivers behind the transition experienced by massive star-forming galaxies during the main epoch of star-formation activity into ``red and dead'' systems \citep[e.g.,][]{rhc_silk98,rhc_springel05,rhc_king15}. Outflows are expected to play (at least) a twofold role in quenching the star formation activity in massive galaxies. First, they have the power to expel the star-forming material in nuclear regions to large distances ($\gtrsim5$~kpc) on short timescales ($\lesssim1$~Gyr) \citep[ejective feedback; e.g., ][]{rhc_feruglio10,rhc_cicone14,rhc_sturm11,rhc_morganti16,rhc_veilleux17}. Second, outflows can transfer enough radiative energy to the circumgalactic medium to drastically reduce the gas accretion rate onto the host galaxy \citep[preventive feedback; e.g., ][]{rhc_dimatteo05,rhc_gabor11,rhc_vdv11}.

To quantify the impact of AGN feedback in galaxy evolution it is important to understand the complex, multi-phase nature of  AGN-driven outflows. This is a challenging task that requires probing winds on a wide range of physical scales ($\sim1~{\rm pc}-10$~kpc), temperatures ($\sim10^2-10^7$~K) and densities ($\sim10^2-10^8$~cm$^{-3}$). The multiple outflow components include (1) very hot, quasi-relativistic winds from accretion disks observed on $\sim$pc scales in X-ray spectra \citep{rhc_nardini15,rhc_tombesi15}, and (2) $\sim$hundred to kilo-parsec scale atomic, molecular, and ionized winds traced by a variety of lines in absorption \citep[e.g., \hi, OH, \naid~D, \feii; ][]{rhc_rupke02,rhc_sturm11,rhc_veilleux13,rhc_kornei12,rhc_gonzalez-alfonso17} and emission \citep[e.g., CO, \cii, \ha, \oiii, etc.;][]{rhc_rupke11,rhc_contursi13,rhc_garcia-burillo14,rhc_woo16,rhc_harrison16,rhc_janssen16,rhc_concas17}. Accurate measurements of the extent, mass and energetics of AGN-driven outflows can only be achieved when more than one gas phase measurement is available, and even then, they are challenging \citep[e.g., Mrk~231; ][]{rhc_feruglio10,rhc_alatalo15,rhc_morganti16}. Part of the problem is that the electron density and the CO-to-H$_2$ factor are required to convert ionized and molecular luminosities into outflow masses, respectively.  These quantities are poorly constrained, and can introduce up to an order of magnitude uncertainty in the derived outflow masses.

Powerful AGN-driven outflows are found to be ubiquitous in {\it typical} massive galaxies at $z\sim2$ \citep[e.g., ][]{rhc_genzel14,rhc_f-s14,rhc_leung17,rhc_f-s18b}, the epoch when both star formation and nuclear activity peaks. These outflows, mostly detected in ionized gas in near-infrared spectroscopic surveys, are fast ($v_{\rm out}\sim1000-2000$~km~s$^{-1}$), can extend on large scales ($R_{\rm out}\sim5-10$~kpc), and have mass loading factors ($\eta\equiv \dot{M}_{\rm out}/{\rm SFR}$) in the $\eta\sim0.1-2$ range \citep[e.g.,][]{rhc_genzel14,rhc_f-s18b}. 

AGN-feedback in the form of outflows is one of the main candidates to explain the observed suppression of star formation activity from nuclear regions outwards in massive, main-sequence galaxies at $z\sim2$ \citep[also referred as {\it inside-out} quenching or growth; ][]{rhc_tacchella15b,rhc_barro16,rhc_nelson16}. However, this would probably require mass loading factors higher than those measured in the ionized phase, which is the reason why it is so important to detect the molecular gas in outflows --most likely the wind phase carrying most of the ejected mass from the host galaxy \cite[e.g., see compilation in][]{rhc_fiore17}.

Wide-band receivers on new or upgraded mm-wave arrays such as ALMA (Atacama Large Millimeter/submillimeter Array) and NOEMA (NOrthern Extended Millimeter Array) opened new windows to detect and characterize the faint (relative to the disk luminosity) molecular outflow signatures in {\it typical} massive galaxies at high $z$. In this paper we present one case where we combine sensitive, high-spatial resolution ALMA and Very Large telescope (VLT) observations to study the molecular and ionized gas phases of the powerful galactic outflow detected in zC400528 at $z\sim2$.

\begin{figure}
\begin{center}
\includegraphics[scale=0.13]{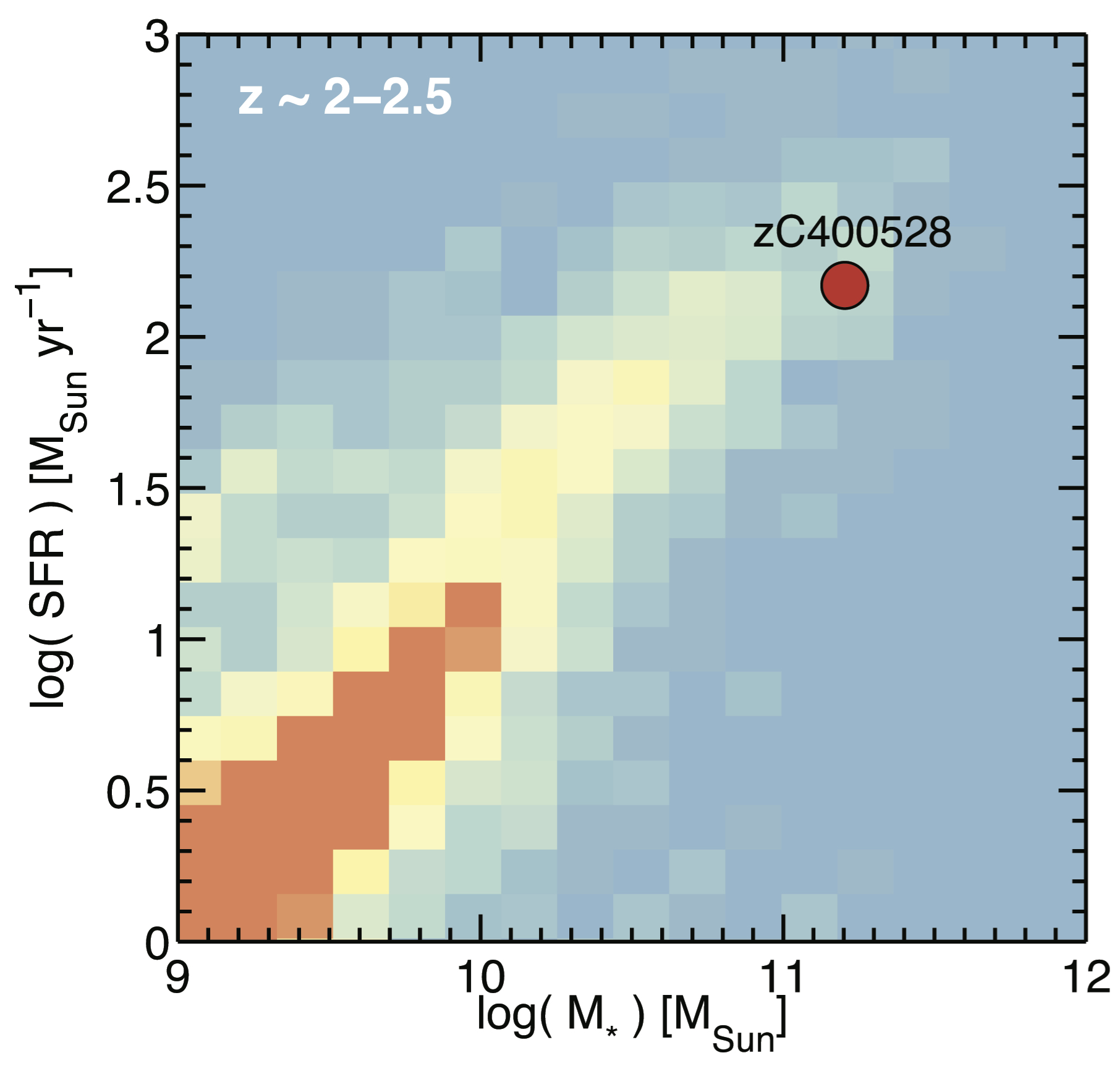}
\caption{Location of zC400528 (red circle) in the stellar mass--star formation rate plane for galaxies in the redshift range $2\leq z \leq2.5$ selected from 3D-HST \citep{rhc_skelton14}. With a stellar mass of $M_{\star}=1.1\times10^{11}~M_{\odot}$ and a $SFR=148$~$M_{\odot}$~yr$^{-1}$, zC400528 lies at the massive end of the main-sequence of star-forming galaxies.}  \label{fig_MS}
\end{center}
\end{figure}

\subsection{zC400528}\label{galaxy_info}

zC400528 (R.A.$=09:59:47.6$, Decl.$=+01:44:19.0$) is a massive ($M_{\star}=1.1\times10^{11}~M_{\odot}$), star-forming (${\rm SFR}=148~M_{\odot}~{\rm yr}^{-1}$\footnote{Position of CO(3-2) peak.}) galaxy at $z=2.3873$ \citep[][]{rhc_f-s18} located at the tip of the main-sequence relation of star-forming galaxies at $z\sim2-2.5$ (see Figure~\ref{fig_MS}). This galaxy has a detected powerful ionized gas outflow that extends for about $\sim5$~kpc \citep[][]{rhc_genzel14, rhc_f-s14}. As discussed in \cite{rhc_f-s14}, there is evidence for a central AGN driving the outflow that includes an observed high nuclear \nii/\ha\ line ratio of 0.75 \citep{rhc_genzel14}, and a detection at 1.4~GHz that implies a ${\rm SFR}\approx790$~M$_{\odot}$~yr$^{-1}$ \citep{rhc_schinnerer10}, exceeding its ${\rm SFR}=148~M_{\odot}~{\rm yr}^{-1}$. The galaxy remains undetected in the {\it Chandra} X-ray data of the COSMOS field \citep{rhc_elvis09}. The flux upper limit implies $L_{\rm X-ray}\lesssim10^{44}$~erg~s$^{-1}$ (assuming $\Gamma=1.9$ and $N_{\rm H}=10^{22}$~cm$^{-2}$), although this upper-limit could be substantially higher in the Compton-thick case.
 
Finally, zC400528 is one of the galaxies in the study of \cite{rhc_tacchella15b} that shows evidence for inside-out quenching on short timescales ($\lesssim1$~Gyr) in the inner $\sim$kiloparsec region (see their Figure~S4, last panel). From a statistical point of view we would also expect that zC400528 will experience star-formation quenching as the fraction of quiescent galaxies with the stellar mass of zC400528 increase from $\sim$20\% at $2<z<2.5$ to $\sim$70\% at $1<z<1.5$ \citep{rhc_muzzin13}.

This paper is organized as follows. In Section~\ref{sec:observations} we describe the observations and the data reduction. In Section~\ref{sec:results} we describe the properties (mass, size, energetics) of the molecular gas in the disk and the outflow. In Section~\ref{sec:comp} we compare the molecular and ionized phases of the outflow. In Section~\ref{sec:escape} we discuss whether the ejected molecular gas can escape the host, the impact of AGN-flickering and the expansion of the outflow, and the effect of AGN-feedback in quenching the star formation activity in the galaxy. Finally, in Section~\ref{sec:conclusions} we present our conclusions.

\begin{figure*}
\begin{center}
\includegraphics[scale=0.23]{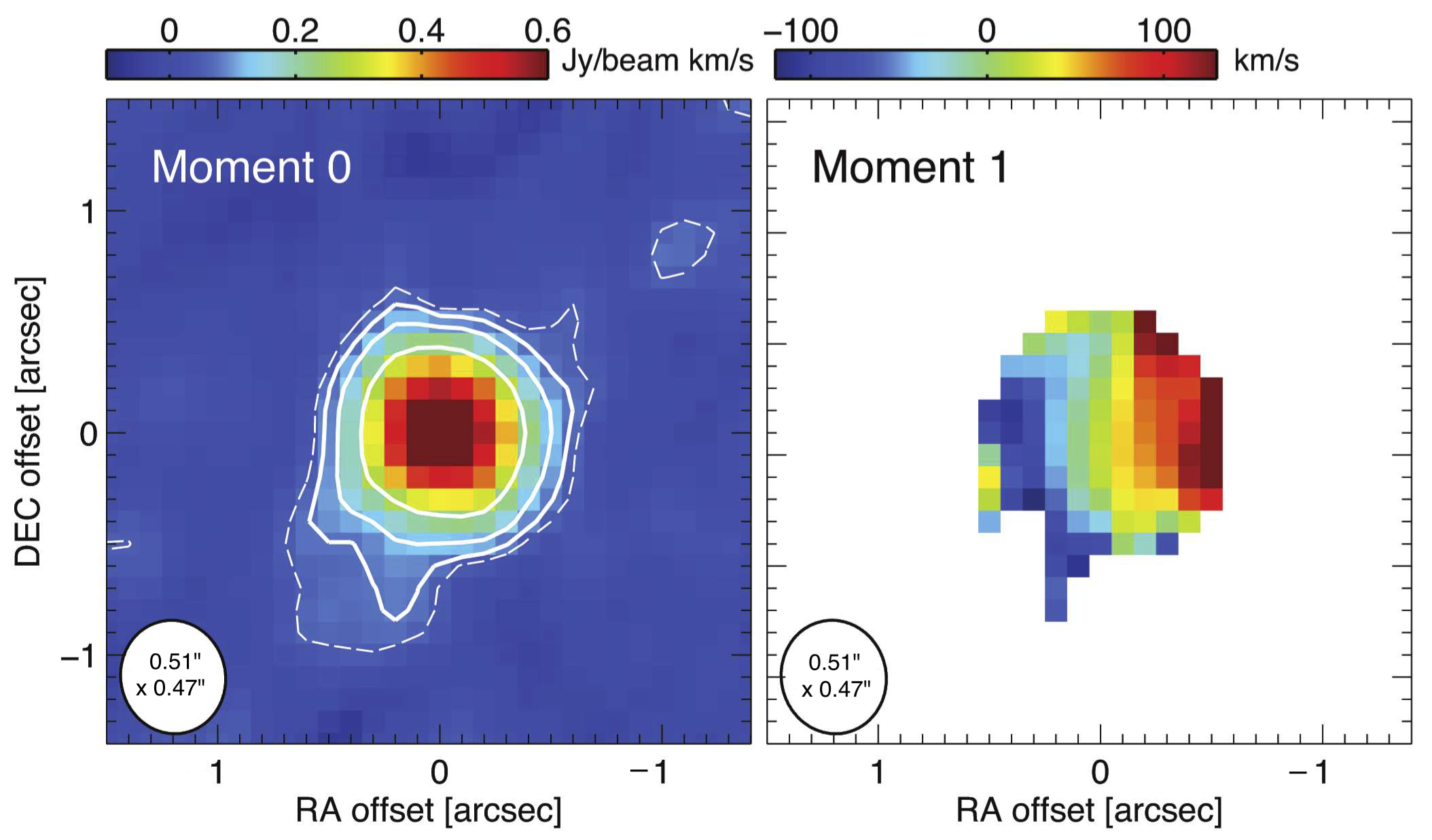}
\caption{ALMA CO(3--2) integrated intensity (left) and velocity (right) maps of zC400528. The contours in the left panel correspond to 2 (dashed), 3, 5, and 10$\sigma$. For the velocity map we only include regions with integrated flux emission higher than 3$\sigma$. The synthesized beam is illustrated in the bottom-left corner. The color bar in the top of each panel indicates the intensity scale.}  \label{fig_moms}
\end{center}
\end{figure*}

\section{Observations and Data Reduction} \label{sec:observations}

The main goal of our observations was to detect and spatially resolve the molecular gas outflow in zC400528 using the redshifted CO(3-2) line emission as a tracer of molecular gas. Our target was first observed by ALMA in Cycle 2 for a total of 0.75~hours (0.6 hours on-source) as part of project 2013.1.00092.S (PI R. Genzel). The observations were carried out in the C34-5/6 configuration, resulting in an angular resolution of $1.1\arcsec \times 0.6\arcsec$ in Band~3. During the next ALMA cycle, our source was observed as part of project 2015.1.00220.S (PI R. Genzel) in a more extended configuration, C40-6, resulting in an angular resolution of $0.4\arcsec \times 0.3\arcsec$ in Band~3. The total observation time was 5.4~hours (4.5~hours on-source). In both cases the spectral setup included one of the spectral windows in Band~3 centered around the redshifted CO(3-2) line emission ($\nu_{\rm obs}=102.004$~GHz).

The data were processed using the Common Astronomy Software Applications package \citep[CASA; ][]{rhc_casa}. The pipeline-calibrated interferometric visibilities delivered by ALMA from both cycles were then combined and imaged at 50~km~s$^{-1}$ resolution using natural weighting. In the combination process we took into account  the different CASA visibility weights applied during Cycle~2 and 3 using the CASA task \textsc{statwt}.\footnote{For more details see: \url{https://casaguides.nrao.edu/index.php/DataWeightsAndCombination}}  The size of the restoring beam of the combined data was $0.51\arcsec\times 0.47\arcsec$ ($4.2\times3.9$~kpc) at a position angle of ${\rm PA=-12.13^{\circ}}$. By combining the data from the more extended and compact array configurations we achieved high enough spatial resolution to resolve the galaxy, and at the same time improve our ability to detect and characterize a potential extended, more diffuse component of the molecular outflow.

All values were primary beam corrected for all quantitative analyses. We reached an rms noise of 0.08~mJy~beam$^{-1}$ in 50~km~s$^{-1}$ channels for CO(3-2). No continuum emission was detected at 100~GHz. 

For a detail description of the observations of the H$\alpha$ and \nii\ lines in zC400528 carried out with the near-IR integral field spectrograph SINFONI \citep{rhc_eisenhauer03} at the Very Large Telescope of the European Southern Observatory (ESO) we refer to \cite{rhc_f-s14}. In summary the galaxy was observed in $K$-band for an on-source time of 4~hr using a natural guide star for the adaptive optics correction. The achieved angular resolution (PSF FWHM) of the final reduced data was 0.15\arcsec.

\section{Results}\label{sec:results}

\subsection{The molecular disk}

Figure~\ref{fig_moms} shows the CO(3-2) integrated intensity and velocity maps of zC400528. We measure an integrated CO(3-2) flux of $F_{\rm CO}=1.11\pm0.05$~Jy~km~s$^{-1}$, which corresponds to a luminosity of $L_{\rm CO}=3.45\times10^{10}~L_{\odot}$. \footnote{This measurement does not include the contribution of the outflow to the CO(3-2) emission. Please see \S~\ref{separation} for a description of how the molecular gas emission of the disk and the outflow was disentangled.} The molecular gas mass ($M_{\rm mol,disk}$), assuming a conversion factor $\alpha_{\rm CO(1-0),T18}=4.3~M_{\odot}~({\rm K~km~s^{-1}~pc^{-2}})^{-1}$ \cite[][]{rhc_tacconi18}\footnote{The $\alpha_{\rm CO}$ conversion function in \cite{rhc_tacconi18} corresponds to the geometric mean between the $\alpha_{\rm CO}$ recipes as a function of metallicity of \cite{rhc_bolatto13} and \cite{rhc_genzel15}.  zC400528 is a massive ($M_{*}=1.1\times10^{11}~M_{\odot}$), main-sequence galaxy at $z=2.38$ that according to the scaling relations in \cite{rhc_tacconi18} has a metallicity near solar ($12+log(\rm O/H)=8.6$) in the \cite{rhc_pp04} scale.} and a velocity-integrated Rayleigh-Jeans brightness temperature line ratio $R_{13}=1.3$ \citep[e.g.,][]{rhc_dannerbauer09,rhc_daddi15,rhc_bolatto15}, is $M_{\rm mol,disk}=1.1\times10^{11}~M_{\odot}$.\footnote{The calculation of the molecular gas mass includes a helium correction factor of 1.36.} This corresponds to a gas to stellar ratio of $\mu_{\rm mol,disk}=M_{\rm mol}/M_{*}\approx1$, which is consistent with typical gas mass fractions observed in massive, star-forming galaxies at $z\sim2$ \citep[e.g.,][]{rhc_tacconi10,rhc_tacconi13,rhc_tacconi18}. 

The integrated velocity map on the right panel of Figure~\ref{fig_moms} shows that the molecular disk is rotating approximately in the east-west direction, consistent with the observed rotation in the ionized gas \citep[][]{rhc_newman13,rhc_f-s18}. 

\begin{figure*}
\begin{center}
\includegraphics[scale=0.235]{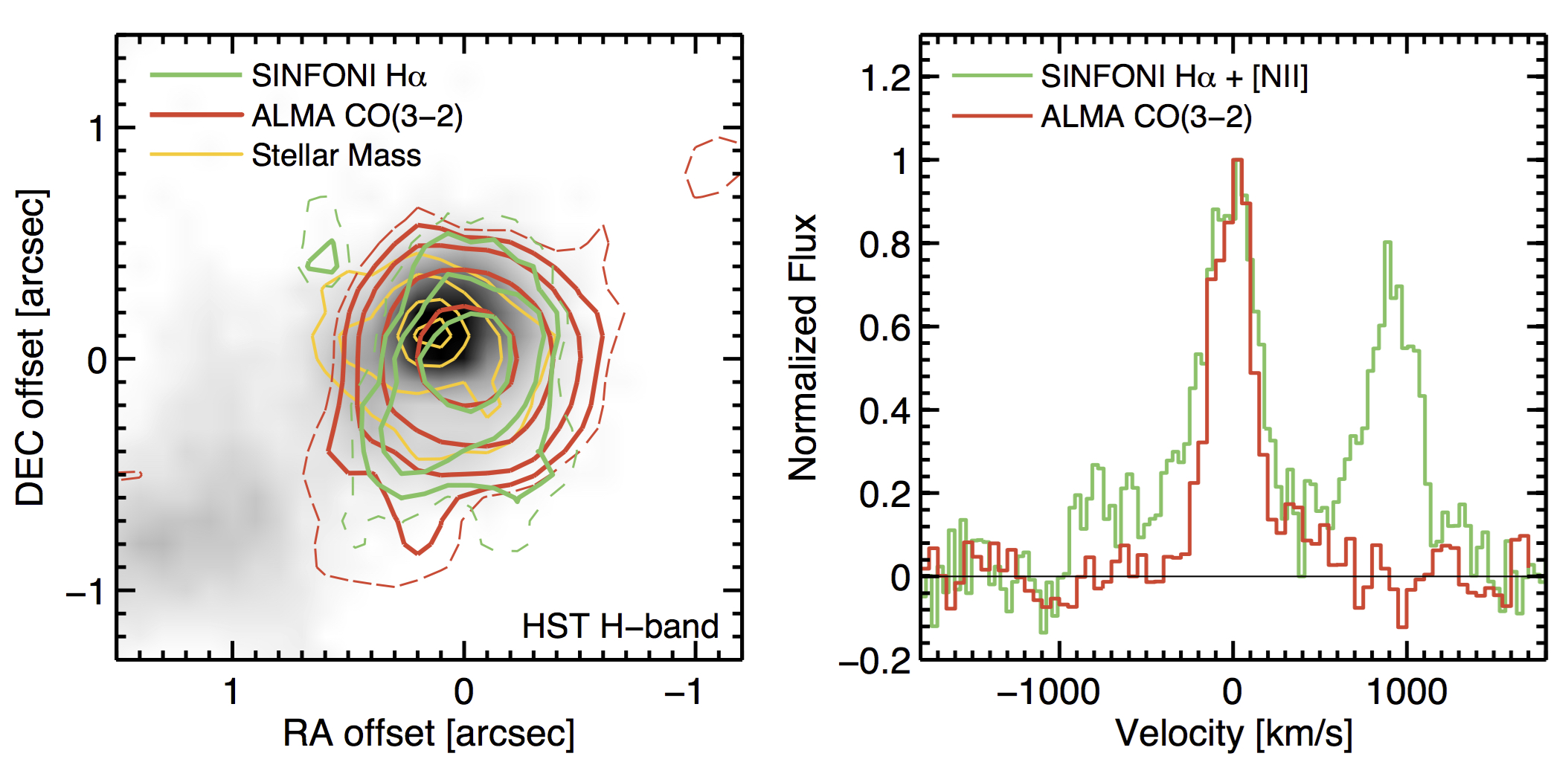}
\caption{{\it (Left)} Contours from ALMA CO(3--2) (red) and SINFONI H$\alpha$ (green) integrated intensity maps, and the stellar mass distribution (yellow) overlaid on a greyscale {\it HST} $H$-band image of zC400528 . For the CO(3-2) and \ha\ data the contours correspond to 2 (dashed), 3, 5, 10, and 20$\sigma$. The contours in stellar mass are 1.5, 3, 6 and $9\times10^{8}$~M$_{\odot}$. We observe that the CO(3--2) emission is offset $\sim0.2\arcsec$ relative to the spatial position of the {\it HST} $H$-band and stellar mass maps. {\it (Right)} ALMA CO(3--2) and SINFONI H$\alpha$+[NII] spectrum within a $0.6\arcsec$-radius circular aperture centered at the peak of emission.  For both ionized and molecular tracers there is evidence for a broad emission component associated with an AGN-driven outflow.} \label{fig_spectra}
\end{center}
\end{figure*}

Figure~\ref{fig_spectra} (left) compares the spatial distribution of the molecular gas, the ionized gas disk traced by H$\alpha$ emission \citep{rhc_f-s14}, and the stellar mass traced by the {\it HST} $H$-band \citep{rhc_tacchella15}. Both molecular and ionized gas components share a similar spatial structure, including a more diffuse component that extends towards the south-east region. The peak of the $H$-band emission is shifted $\sim0.2\arcsec$ north-east with respect to the peak of the ALMA CO(3-2) emission. The astrometric precision of our ALMA data is $\sim0.025\arcsec$,\footnote{According to the ALMA Technical Handbook, Chapter 10.6.6, Astrometric Observations} which is at least a factor of $\sim10$ better than the astrometrical accuracy of the {\it HST} data. Therefore, it is not possible to determine whether the spatial offset is due to physical reasons or astrometric errors.

The right panel of Figure~\ref{fig_spectra} shows the CO(3-2) and H$\alpha$+\nii\ spectrum from within a $0.6\arcsec$-radius circular aperture centered at the peak of emission. The \nii/\ha\ line peak ratio is 0.8 \citep[see also][]{rhc_genzel14} which is characteristic of galaxies where the AGN contributes to the ionizing radiation or shocks affect the ionization balance \citep[e.g.,][]{rhc_kewley13,rhc_newman14}. As described in detail in \cite{rhc_genzel14} and \cite{rhc_f-s14}, an ionized outflow is detected as a strong nuclear broad component in the \ha\ spectrum (detected out to velocities $\sim\pm1000$~km~s$^{-1}$ relative to the systemic velocity). In the case of the CO(3-2) spectrum, we observe a high-velocity wing on the red side ([+300,+500]~km~s$^{-1}$) that is likely the molecular component of the outflow. We analyze and interpret this high-velocity feature in the CO(3-2) spectrum in the next Section.  

\begin{figure*}[ht!]
\begin{center}
\includegraphics[scale=0.22]{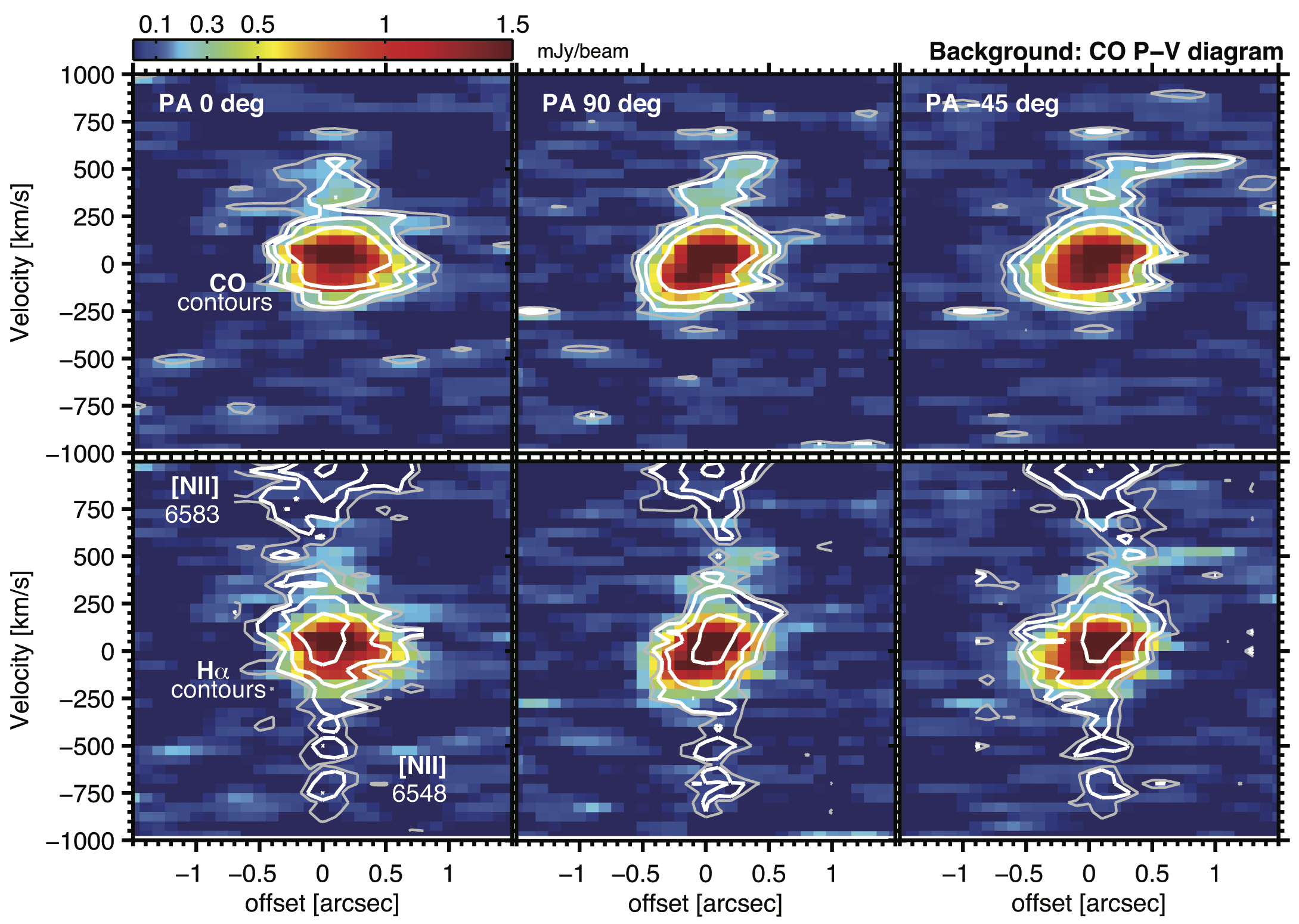}
\caption{({\it Top panels}) Position-velocity diagram of zC400528 for the ALMA CO(3--2) data (background and contours) taken at a position angle of 0$^{\circ}$ (left), 90$^{\circ}$ (middle), and $-45^{\circ}$ (right), through the CO(3--2) emission peak. The contours show the 2 (gray), 3, 6, and 10$\sigma$ (white) levels of emission. ({\it Bottom panels}) Similar to the upper panels, but this time the contours are based on the position-velocity diagram for the SINFONI H$\alpha$+[NII] data. High-velocity gas at $v\gtrsim+300$~km~s$^{-1}$ is seen in both molecular and ionized emission.} \label{fig_pv}
\end{center}
\end{figure*}

\begin{figure*}
\begin{center}
\includegraphics[scale=0.25]{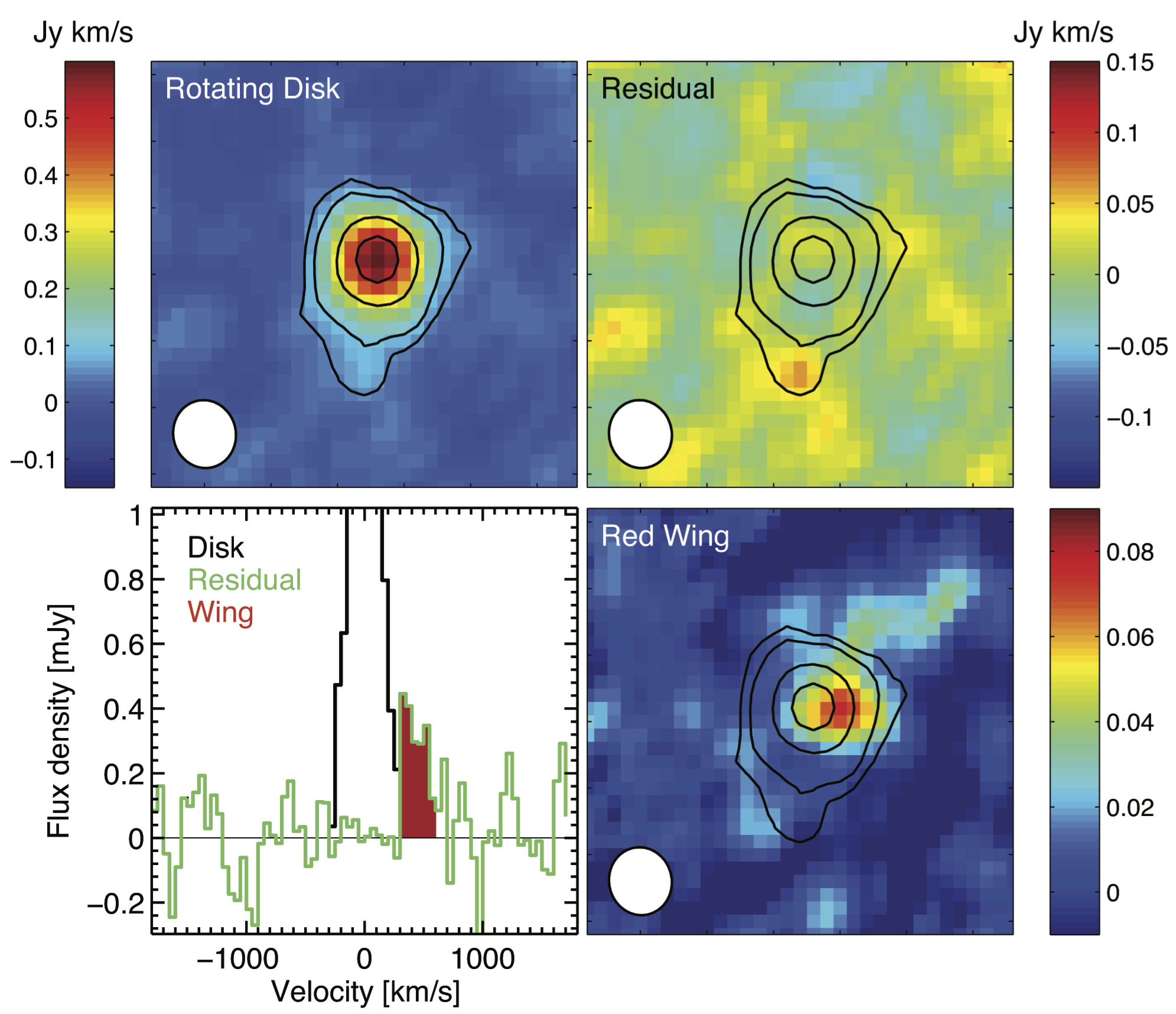}
\caption{Analysis of the kinematic components of zC400528 based on a two-dimensional Gaussian fit to the CO(3--2) emission in the 50~km~s$^{-1}$ channels. {\it (Top panels)} Integrated intensity maps in the [$-250$, 250]~km~s$^{-1}$ range for the rotating disk component (left) and the residual component after subtracting the rotating disk (right). The synthesized beam is illustrated in the bottom-left corner ($0.51\arcsec\times0.47\arcsec$). The color bar indicates the intensity scale. {\it (Bottom-left panel)} CO(3-2) spectrum of the rotating (disk) and residual (green) components within a 0.5\arcsec-radius circular aperture centered 0.2\arcsec west of the peak of CO emission in the rotating disk. In red we show what we identify as the red-wing outflow material. {\it (Bottom-right panel)} Intensity map of the red wing component integrated  between +300 and +650 km~s$^{-1}$. The contours in each map represent the velocity-integrated CO(3-2) emission of the rotating disk. The separation between tick marks is 1\arcsec}\label{model}
\end{center}
\end{figure*}

\begin{figure*}
\begin{center}
\includegraphics[scale=0.24]{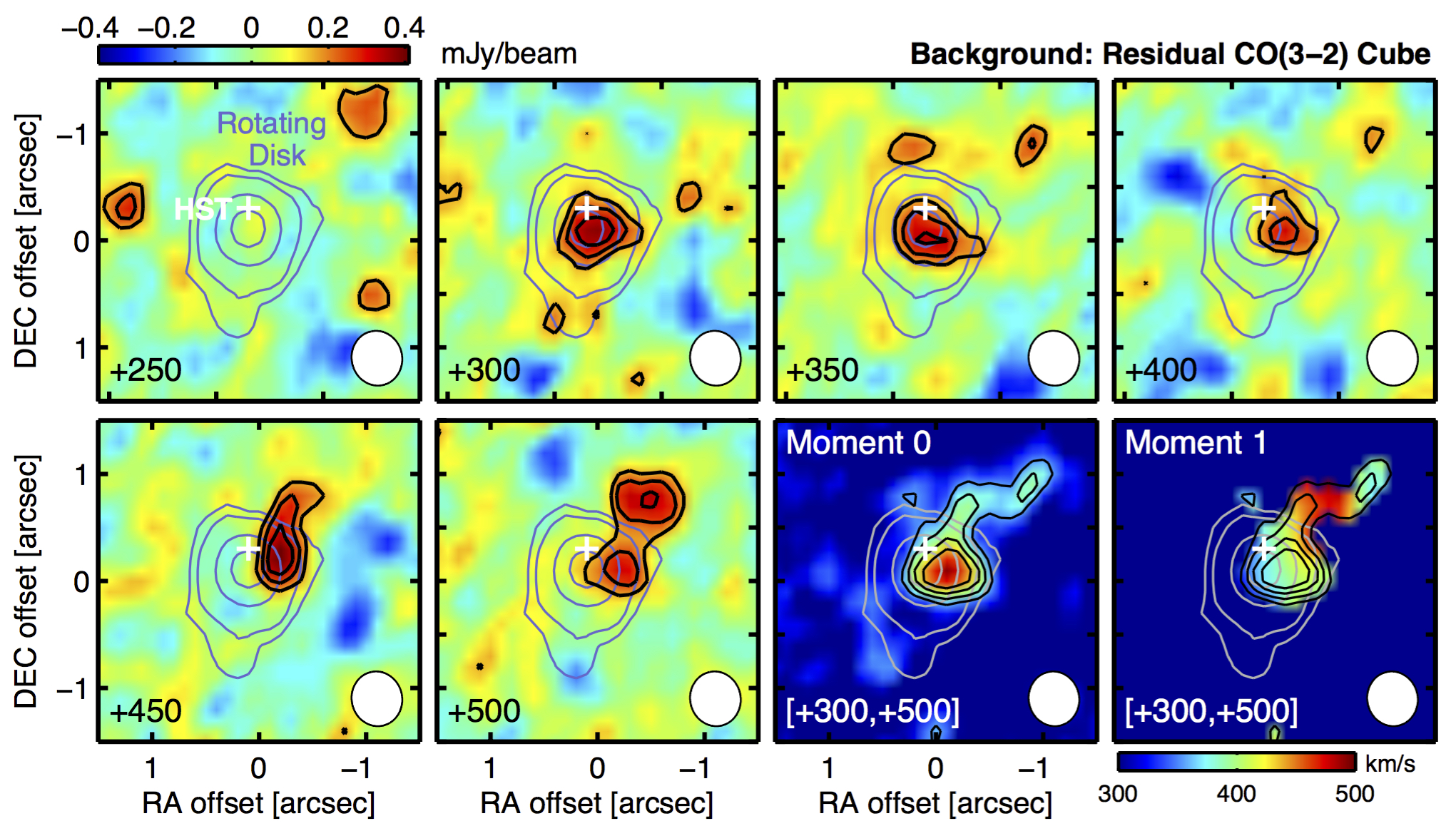}
\caption{Channel maps showing the CO(3--2) emission of zC400528 in the residual cube (i.e., after subtracting the rotating disk model) in the [+250,+500]~km~s$^{-1}$ range. The purple contours show the spatial distribution of the integrated CO(3-2) emission of the rotating disk, and the white cross indicates the position of the peak of emission in the {\it HST} {\it H}-band image. The last two panels show the integrated intensity and velocity maps of the outflow in the [+300,+500]~km~s$^{-1}$ range. The contours start at 2$\sigma$ and have increments of 1$\sigma$. The synthesized beam is shown in the bottom right corner.}\label{vel_chan_map}
\end{center}
\end{figure*}

\subsection{The molecular outflow}

As a first step to study the high-velocity components in the ionized and molecular spectrum of zC400528, we make position-velocity (P-V) diagrams extracted along the vertical (south-north), horizontal (west-east), and $-45^{\circ}$ (south-east to north-west) directions through the emission peak. The top panel in Figure~\ref{fig_pv} shows the results for the CO(3-2) data. We identify the rotation pattern of the galaxy in the $[-250,+250]$~km~s$^{-1}$ velocity range, and a high-velocity  ($v\gtrsim+300$~km~s$^{-1}$) component along all the examined directions, and that preferentially extends along the north-west direction for about $\sim1$\arcsec  (which corresponds to a projected distance of $\approx8.4$~kpc). The contours in the bottom panels of Figure~\ref{fig_pv} show the P-V curves for the \ha\ and \nii\ lines (the systemic velocity is set to match the \ha\ line central velocity) plotted on top the CO(3-2) P-V diagram. On the receding side, the outflow is detected both in ionized and molecular gas, with the latter showing a more extended spatial distribution. On the other hand, there is no molecular counterpart to the high-velocity ionized gas observed at velocities $v\lesssim-300$~km~s$^{-1}$.

\subsubsection{Disentangling the molecular emission from disk and outflow}\label{separation}

To quantify the total amount of molecular gas in the outflow it is necessary to identify and remove the CO(3-2) line emission associated with the rotating disk of the galaxy. One common approach is to fit two Gaussian profiles to the spectrum where a narrow component centered at the systemic velocity is used to characterize the rotating disk, and a broad component to characterize the outflow. The flux in the broad component, however, only represents an upper limit to the actual flux from the outflowing material since it may include CO line emission at low velocities that is associated with the rotating disk. Given that we have sufficient spatial resolution and signal-to-noise (S/N) in our ALMA CO(3-2) data, here we try a different strategy to attempt to remove the low-velocity gas material.

First, we use the CASA task \textsc{imfit} to fit in each 50~km~s$^{-1}$ channel of the CO(3-2) cube a two-dimensional Gaussian to the image in the region where the source is detected. We then use the resulting fits to make a source model where the Gaussian position, orientation, and intensity changes as a function of velocity. The last step in our procedure is to remove from each slice the source model, resulting in a residual cube where the high-velocity outflowing gas is now disentangled from the emission associated with the rotating disk. A similar approach was used by \cite{rhc_veilleux17} to study the molecular outflow in a luminous infrared galaxy using CO(1-0) observations.

The first two panels in Figure~\ref{model} show the integrated intensity map calculated on the [$-$250,+250]~km~s$^{-1}$ range of the original cube (left), and the cube resulting after we subtract the rotating disk model (i.e., the residual cube; right). The model does a good job at removing the CO(3-2) line emission associated with the rotating disk, and we only observe a relatively high residual at the level of $\sim$0.06~Jy~km~s$^{-1}$ in the more diffuse and extended south-east component. The third panel in Figure~\ref{model} shows the CO(3-2) spectrum integrated over a circular aperture with a radius of 0.5\arcsec. The black and green lines correspond to the spectrum extracted in the original cube and the residual cube, respectively. The integrated signal of the high-velocity red wing emission in the residual cube is clear, and there is no evidence for a molecular outflow component on the approaching side of the spectrum. The last panel of Figure~\ref{model} shows the integrated intensity map of the residual cube in the velocity range where we identified the red wing emission ([+300,+700]~km~s$^{-1}$). The outflow emission peaks about 0.2\arcsec\ ($\sim2$~kpc) west of the molecular disk center, and there is a tail of outflowing gas line emission that extends beyond the molecular disk towards the north-west for about 1\arcsec, which corresponds to a projected distance of $\sim8$~kpc.

\subsubsection{The structure of the molecular outflow}\label{section:structure}

A more detailed view of the molecular outflow components as a function of velocity is shown in the velocity channel map of the residual cube in Figure~\ref{vel_chan_map}. For reference, overplotted on each panel are the contours from the velocity integrated CO(3-2) emission of the rotating disk. The first significant outflow structure appears at $v=+300$~km~s$^{-1}$ and is spatially coincident with the central region of the molecular gas disk. Between $v=+350$ and +400~km~s$^{-1}$ the main component of the outflow moves towards the west about 0.2\arcsec\ --which corresponds to a projected distance of $\sim2$~kpc. At $v=+450$~km~s$^{-1}$ the main outflow component starts to extend towards the north and continues until $v=+500$~km~s$^{-1}$, which is the last channel at which we detect CO(3-2) line emission. 

The last two panels in Figure~\ref{vel_chan_map} show the integrated intensity and velocity maps of the molecular outflow detected in the [+300,+500]~km~s$^{-1}$ range. The bulk of the line emission associated with the outflow is shifted about 0.2\arcsec\ (projected distance of $\sim2$~kpc) west of the center of the molecular disk and extends to the north for approximately 0.7\arcsec\ (projected distance of $\sim6$~kpc), where it shifts to the west in what seems to appear a second outflow component --still connected to the main component-- at about 1.2\arcsec\ (projected distance of $\sim10$~kpc) from the center. One could suspect that this spatially extended feature of the outflow is in reality a tidal tail product of the interaction with a lower mass companion, but visual inspection of a deep {\it HST} F160W image of the field reveals no stellar emission associated to a close neighbor in the north-east quadrant.

We observe a velocity gradient in the outflow: the gas velocity increases from $\sim+350$ to $\sim+500$~km~s$^{-1}$ as the outflow extends north of the nucleus. This could be indicative of an accelerating continuous outflow. Another possibility could be that the ejection of molecular gas triggered during the last outflow burst had a distribution of velocities, so the faster ejecta components traveled further away than the slower components. The velocity of the most distant outflow component in the north-west has a velocity of $\sim+350$~km~s$^{-1}$, a lower velocity that could be the result of gravitational pull. 

In summary, perhaps the most striking characteristic of the molecular outflow in zC400528 is its asymmetry: we only detect the receding component of the wind. In the local universe is not uncommon to observe asymmetric outflows. For example, \cite{rhc_p-s18} find that in four out of five spatially resolved molecular gas outflows in ULIRGs the receding component of the wind is stronger than the approaching one. This is also true for the molecular outflow in Mrk~231 \citep{rhc_cicone12,rhc_feruglio15} and the western nucleus of Arp~220 \citep{rhc_barcos18}. High-resolution simulations of $z\sim2$ isolated disks also find that AGN-driven outflows are typically unipolar as a result of dense cloud structures in the vicinity of the black hole blocking the expansion in one direction \citep{rhc_gabor14}. Another possibility is that the molecular outflow on the approaching side experienced a change of phase, similar to what is observed in M82 where the dominant phase of the wind transitions from molecular to atomic at about one kiloparsec distance from the disk \citep[][]{rhc_leroy15}.

\begin{figure*}
\begin{center}
\includegraphics[scale=0.14]{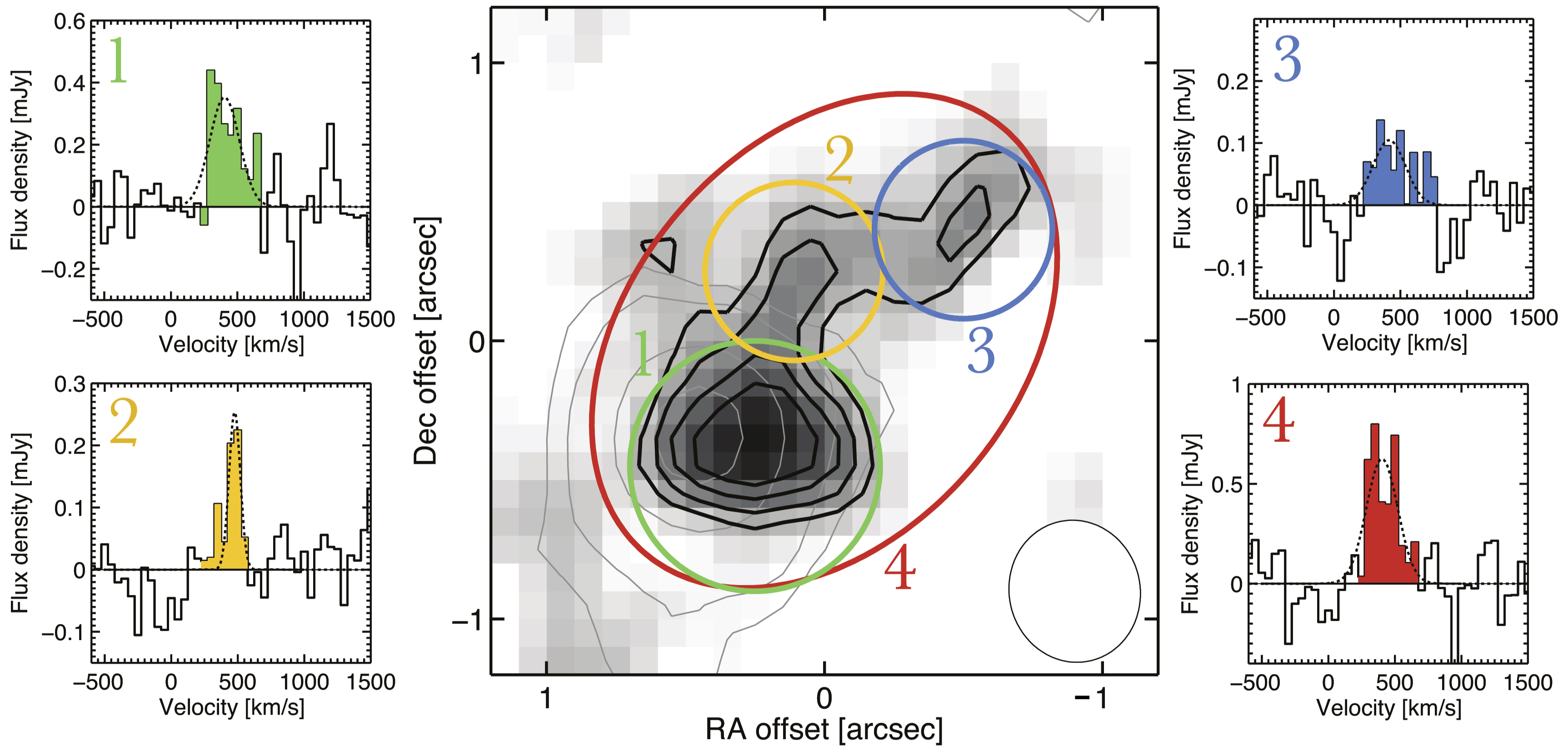}
\caption{(Central panel) Integrated intensity CO(3-2) line emission map of the molecular outflow in zC400528 from the source-model-subtracted cube (see \S\ref{separation}. The side panels show the integrated spectrum extracted from the apertures centered at the different outflow components: (Aperture 1) The central component; (Aperture 2) the tail extending north of the molecular disk; (Aperture 3) the component that could be the result of a previous outflow episode product of the AGN variability; (Aperture 4) all the outflow components. The flux densities measured within each aperture are listed in Table~\ref{table:flux}. The synthesized beam is shown in the bottom right corner.
} \label{flux}
\end{center}
\end{figure*}

\subsubsection{The molecular gas mass in the outflow}

For each outflow component, we measure CO(3-2) flux densities with Gaussian fits to the spectra in the residual cube extracted from the circular and elliptical apertures shown in Figure~\ref{flux}. The flux densities are listed in Table~\ref{table:flux}.

How to convert these flux densities associated with the outflow into molecular gas masses is still an open question. The CO-to-H$_{2}$ conversion factor depends mainly on the metallicity and the surface density of the gas \citep[e.g.,][]{rhc_bolatto13}. In the case of the CO gas in the outflow, in the literature there are three CO-to-H$_{2}$ conversion factors commonly used: (1) a Galactic conversion factor of $\alpha_{\rm CO,MW}=4.3~M_{\odot}~({\rm K~km~s^{-1}~pc^{-2}})^{-1}$ \citep{rhc_bolatto13}; (2) a ULIRG-like conversion factor of $\alpha_{\rm CO,ULIRG}=0.8~M_{\odot}~({\rm K~km~s^{-1}~pc^{-2}})^{-1}$ \citep[e.g.,][]{rhc_cicone14,rhc_feruglio15,rhc_veilleux17}, and (3) an optically-thin conversion factor of $\alpha_{\rm CO,thin}=0.34~M_{\odot}~({\rm K~km~s^{-1}~pc^{-2}})^{-1}$ \citep[e.g.,][]{rhc_bolatto13b,rhc_richings18}.  Assuming that the molecular gas in outflows is optically thin --as observed for example in the jet accelerated wind of IC~5063 \citep{rhc_dasyra16}--, would yield the most conservative (or ``minimum'') estimate of the molecular gas masses.

In this paper we adopt the ULIRG-like conversion factor ($\alpha_{\rm CO,ULIRG}$), which seems to be a good compromise given the range of conversion factors available. In addition, the detection of dense molecular gas entrained in the outflow of starbursts and LIRGs \citep[e.g.,][]{rhc_aalto12,rhc_aalto15,rhc_walter17} argues in favor a conversion factor higher than $\alpha_{\rm CO,thin}$.  In the main streamer of NGC~253, for example, the outflowing gas is not optically thin, as a the observed CO(2-1) to CO(1-0) brightness temperature line ratio is about unity (Zschaechner et al., submitted to ApJ). We also expect that the molecular gas in the outflow is likely not cold but warm, because it has been shocked at some level and/or it is immersed in hot gas and subjected to a substantial external radiation field. In that case the physical conditions of the molecular gas in the wind  better resemble those found in the ISM of (U)LIRGs rather than the Milky Way.

\begin{figure}
\begin{center}
\includegraphics[scale=0.12]{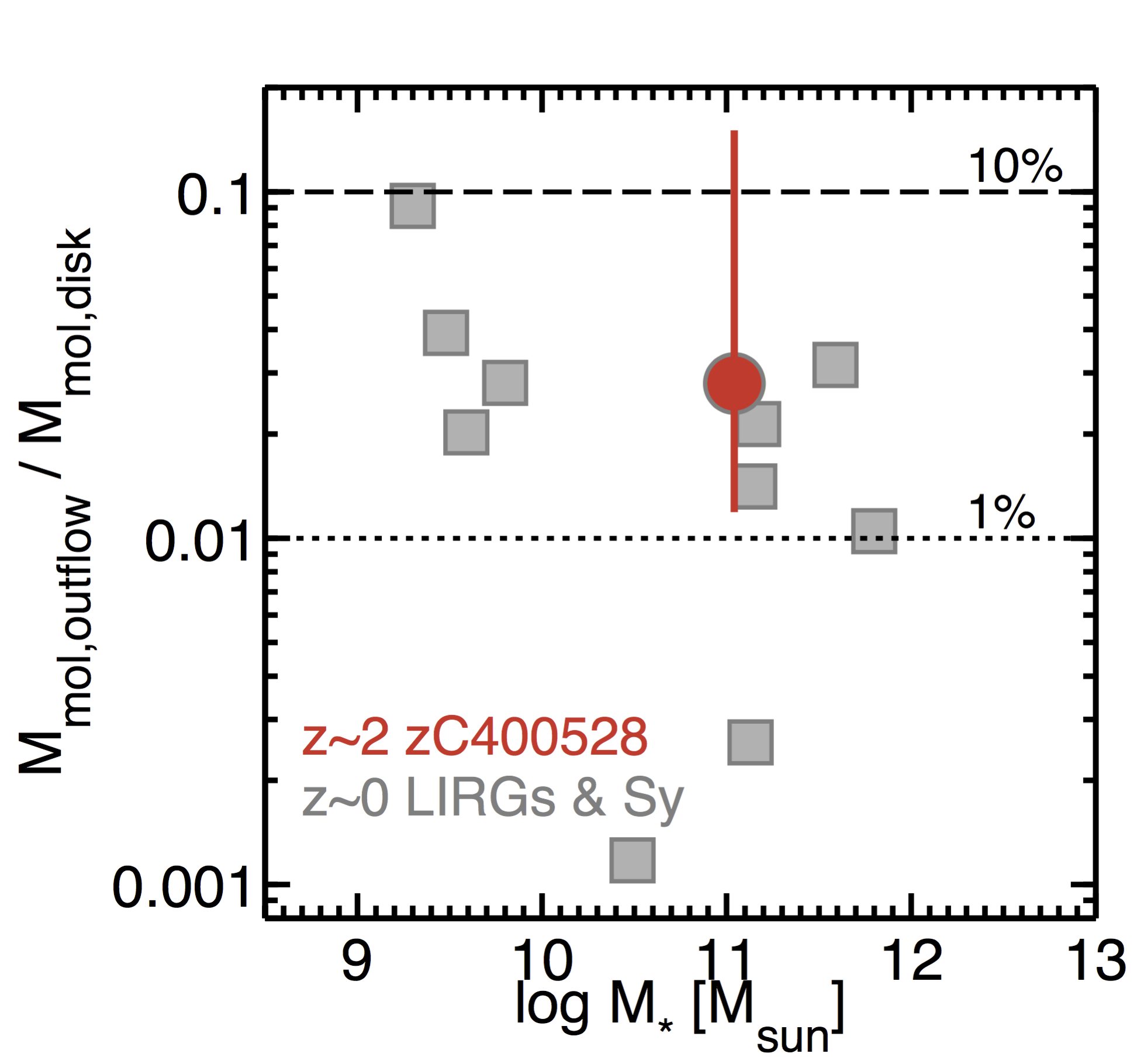}
\caption{Ratio between the molecular gas mass of the AGN-driven outflow and the galaxy as a function of stellar mass for local (U)LIRGs and Seyferts \citep[][and references therein]{rhc_fiore17} and our target, zC400528. In all cases we assume an $\alpha_{\rm CO,ULIRG}$ conversion factor to estimate the molecular gas mass in the outflow. The vertical red line associated with the molecular gas fraction in zC400528 show how the value would change if we assume an optically thin or MW CO-to-H$_{2}$ conversion factor instead of $\alpha_{\rm CO,ULIRG}$} \label{frac}
\end{center}
\end{figure}

The molecular gas masses for each of the outflow components are listed in Table~\ref{table:flux}. The total molecular gas mass in the outflow of zC400528 is

\begin{equation*}
M_{\rm out,mol}=3.36\times10^{9}~M_{\odot}\times\bigg(\frac{\alpha_{\rm CO}}{\alpha_{\rm CO,ULIRG}}\bigg),
\end{equation*}

\noindent which corresponds to $\sim3\%$ of the molecular gas mass in the disk. As Figure~\ref{frac} shows, this fraction is similar to those measured in local (U)LIRG and Seyfert galaxies of similar stellar mass \citep[][and references therein]{rhc_fiore17}\footnote{Note that for the bodies of starburst and (U)LIRGs drawn from the literature that are outliers in the local main-sequence relation ($\Delta_{\rm MS}\gtrsim1$~dex) we use the CO-to-H$_{2}$ conversion factor from the scaling relation in \cite{rhc_genzel15} to calculate their molecular gas masses. This conversion factor is a factor $\sim2$ lower for starbursts and (U)LIRGs galaxies depending on their specific star formation rate, and similar to that in \cite{rhc_tacconi18} for main-sequence galaxies.}.

\begin{deluxetable}{cccccc}
\tabletypesize{\footnotesize} 
\tablecaption{Molecular outflow mass in zC400528\label{table:flux}}
\tablehead{
\colhead{Component} & \colhead{Integrated Flux} & \colhead{$M_{\rm out,mol}$\tablenotemark{a}}  \\
\colhead{\#} & \colhead{[Jy~km~s$^{-1}$]} & [$10^{9}~M_{\odot}$]}  
\startdata
1 & $0.098\pm0.016$ & 1.81 $(0.77-9.72)$ \\
2 & $0.029\pm0.007$ & 0.50 $(0.21-2.68)$ \\ 
3 & $0.032\pm0.007$ & 0.59 $(0.25-3.17)$ \\ 
\hline
4 & $0.184\pm0.029$ & 3.36 $(1.42-18.06)$ \\ 
\enddata
\tablenotetext{a}{Molecular gas masses calculated using an $\alpha_{\rm CO,ULIRG}$ conversion factor. The values in parenthesis correspond to the molecular gas masses we would obtain if we apply a $\alpha_{\rm CO,thin}$ or $\alpha_{\rm CO,MW}$ conversion factor, respectively.}
\end{deluxetable}

\begin{figure}
\begin{center}
\includegraphics[scale=0.115]{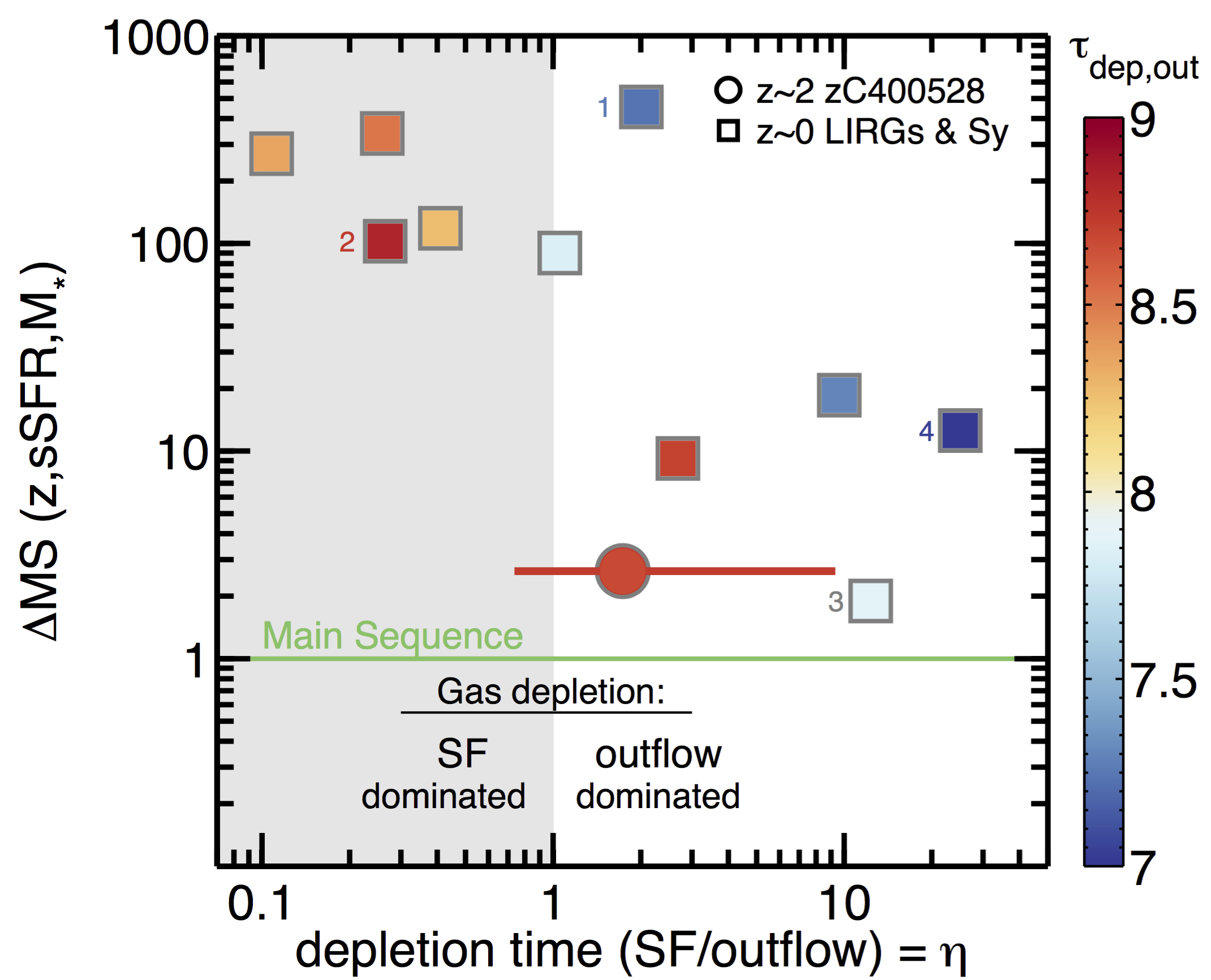}
\caption{Separation from the main-sequence of star-forming galaxies --after removing the dependence with $z$ using \cite{rhc_tacconi18} scaling relations--, versus the ratio between the star-formation and the outflow depletion timescales, or equivalently, the mass loading factor ($\eta=\dot{M}_{\rm out}/{\rm SFR}$). In color we show the logarithm of the outflow depletion timescale in Gyr units. zC400528 is shown as a circle, and local LIRGs and Seyfert galaxies are plotted as squares. The unshaded area shows where $\tau_{\rm dep,out}<\tau_{\rm dep,SF}$, i.e., where the outflow in the system is powerful enough to deplete the galaxy molecular gas reservoir in a timescale shorter than that needed to exhaust it by star formation. For reference, galaxies 1, 2, 3, and 4 are Mrk~231, NGC~6240, NGC~1433, and NGC~1068.} \label{dMS}
\end{center}
\end{figure}

\subsubsection{Molecular mass outflow rate}\label{section:Mdot}

There are two common approaches to measure outflow mass rates in galaxies. There are often referred as {\it instantaneous} (maximum) and {\it average} (minimum), and a detailed description of the assumptions that go into the calculations can be found in \cite{rhc_rupke05}. In the {\it instantaneous} approach the outflow mass rate ($dM_{\rm out}/dt$) is calculated as the product of the outflow gas mass and the timescale taken by the gas to cross the thickness of the outflowing shell, i.e., 

\begin{equation}
\dot{M}_{\rm out}^{\rm inst} = M_{\rm out}\times\frac{v_{\rm out}}{\Delta R},
\end{equation}

\noindent where $v_{\rm out}$ is the velocity of the outflow and $\Delta R$ ($=R_{\rm out}-R_{\rm int}$) is the thickness of the outflow shell. This approach has been used in a number of studies including \cite{rhc_sturm11} and \cite{rhc_gonzalez-alfonso14}.

In the case of the {\it average} approach, the assumption is that the outflowing gas extends to $r = 0$, so the outflow mass rate ($\dot{M}_{\rm out}$) is given by the outflow gas mass time-averaged over the flow timescale, i.e., 

\begin{equation} \label{eq:avg}
{\dot M}_{\rm out}^{\rm avg} = M_{\rm out}\times \frac{v_{\rm out}}{R_{\rm out}}.
\end{equation}

\noindent If the emitting volume (spherical or multi-conical) is filled with uniform density, then the mass outflow rate in Eq.~\ref{eq:avg} should be a factor of 3 higher \citep[e.g.,][]{rhc_feruglio10,rhc_maiolino12,rhc_r-z13,rhc_fiore17}. 

In this paper we use the {\it average} approach, which represents a more conservative way to characterize the outflow than the {\it instantaneous} method as $\dot{M}_{\rm out}^{\rm avg}$ is smaller than $\dot{M}_{\rm out}^{\rm inst}$ by a factor $(R_{\rm out}/\Delta R)$. Following this, the first step is to calculate the flow timescale, $t_{\rm out} = R_{\rm out}/v_{\rm out}$. As described in \S\ref{section:structure}, zC400528 has three main molecular outflow components. If we estimate the flow timescale based on the maximal extension of the outflow, then $t_{\rm out,mol}\approx8.5~{\rm kpc}/400~{\rm km~s}^{-1}\approx2\times10^7$~yr. A more representative or characteristic timescale would result from considering only the main outflow component (which corresponds to regions 1 and 2 in Figure~\ref{flux}). This component encompasses  $\sim80$\% of the outflow total mass, has a size of $R_{\rm out,mol}=0.5\arcsec$ ($\approx4.2$~kpc; calculated from a two-dimensional Gaussian fit and deconvolved from the beam), and shows a positive velocity gradient in the [+350,+500]~km~s$^{-1}$ range. From these values the resulting flow timescale is $t_{\rm out,mol}\approx4.2~{\rm kpc}/450~{\rm km~s}^{-1}\approx9\times10^6$~yr, which yields a mass outflow rate of:

\begin{equation*}
\begin{split}
\dot{M}_{\rm out,mol}^{\rm avg}\approx256~M_{\odot}~{\rm yr}^{-1}\times\bigg(\frac{\alpha_{\rm CO,out}}{\alpha_{\rm CO,ULIRG}}\bigg)\\
\times\bigg(\frac{v_{\rm out}}{450~{\rm km}~{\rm s}^{-1}}\bigg)\times\bigg(\frac{4.2~{\rm kpc}}{R_{\rm out}}\bigg)
\end{split}
\end{equation*}

\noindent This translates into an outflow depletion time ($\tau_{\rm dep,out}= M_{\rm mol,disk}/\dot{M}_{\rm out,mol}^{\rm avg}$) of

\begin{equation*}
\begin{split}
\tau_{\rm dep,out}\approx4.7\times10^8~{\rm yr} \times \bigg(\frac{\alpha_{\rm CO,disk}}{\alpha_{\rm CO,T18}}\bigg) 
\times \bigg(\frac{\alpha_{\rm CO,ULIRG}}{\alpha_{\rm CO,out}}\bigg) \\
\times \bigg(\frac{450~{\rm km}~{\rm s}^{-1}}{v_{\rm out}}\bigg)\times
\bigg(\frac{R_{\rm out}}{4.2~{\rm kpc}}\bigg).
\end{split}
\end{equation*}

\noindent Compared to the star-formation depletion timescale ($\tau_{\rm dep,SF}=M_{\rm mol,disk}/{\rm SFR}$), $\tau_{\rm dep,out}$ is a factor of $\sim2$ shorter -- or equivalently, the mass loading factor in the molecular phase ($\eta=\dot{M}_{\rm out,mol}/{\rm SFR}$) is $\sim2$.

A more common approach used in outflow studies to measure mass loss rates consist in fitting the system line emission using two Gaussian components -- a narrow component for the disk and a broad component for the outflow. The ``maximum'' projected velocity of the outflow component is then defined as the centroid velocity plus one-half the velocity width \citep[e.g.,][]{rhc_rupke05,rhc_veilleux05}. Following this method we measure $M_{\rm out}=3.6\times10^{9}$~M$_{\odot}$ and $v_{\rm out,max}=590$~km~s$^{-1}$. If we consider these values and the maximum projected extension of the outflow, $R_{\rm max}\approx8.5$~kpc, we calculate a mass loss rate of $\dot{M}_{\rm out}=180$~M$_{\odot}$~yr$^{-1}$. This result is consistent, within the uncertainties involved in the calculation, to that we derived using the method adopted in this work described in \S\ref{separation}. 

So far we have ignored the effect that the inclination of the outflow has on the determination of the mass outflow rate, and consequently, the outflow depletion timescale. The reason is that we do not have enough information to constrain the geometry of the outflow. If $\theta$ corresponds to the inclination angle of the outflow with respect to the line-of-sight, then the deprojected outflow size is $R_{\rm out}=R_{\rm out,proj}/{\rm sin}(\theta)$ and the deprojected outflow velocity is $v_{\rm out}=v_{\rm out,proj}/{\rm cos}(\theta)$. This implies that the inclination corrected mass outflow rate is proportional to ${\rm tan}(\theta)$. zC400528 has an inclination of $i\approx37^{\circ}$ \citep[${\rm sin}(i)=0.61$;][]{rhc_f-s18}. If we assume that the outflow is perpendicular to the disk, then the deprojection effects in velocity and radius almost cancel out and the correction is small and of the order of $\sim0.75$.

Figure~\ref{dMS} compares the outflow and star formation depletion timescales of galaxies as a function of their separation from the main-sequence (after removing the dependence with $z$ using \cite{rhc_tacconi18} scaling relations). The plot includes zC400528 and a compilation of local (U)LIRG and Seyfert galaxies \citep[][and reference therein]{rhc_fiore17}. We observe that local systems where the depletion of the molecular gas is dominated by star formation lie at least $\sim1.5$~dex above the main-sequence. On the other hand, galaxies where the depletion of molecular gas is dominated by outflows are located within $\sim1$~dex of the main-sequence, with the exception of the above outlier Mrk~231 \citep{rhc_feruglio10}. Similar to Mrk~231, our target galaxy zC400528 is exhausting its nuclear molecular gas reservoir via the outflow about twice as fast as due to its star formation activity. Note, however, that in contrast to Mrk~231 zC400528 is not a $\times100$ outlier but only +0.4~dex above the main-sequence of star-forming galaxies at $z\approx2$. We discuss in more detail the effect that the powerful molecular outflow detected in zC400528 may have on quenching its star formation in Section~\ref{sec:quenching}.

\begin{figure}
\begin{center}
\includegraphics[scale=0.23]{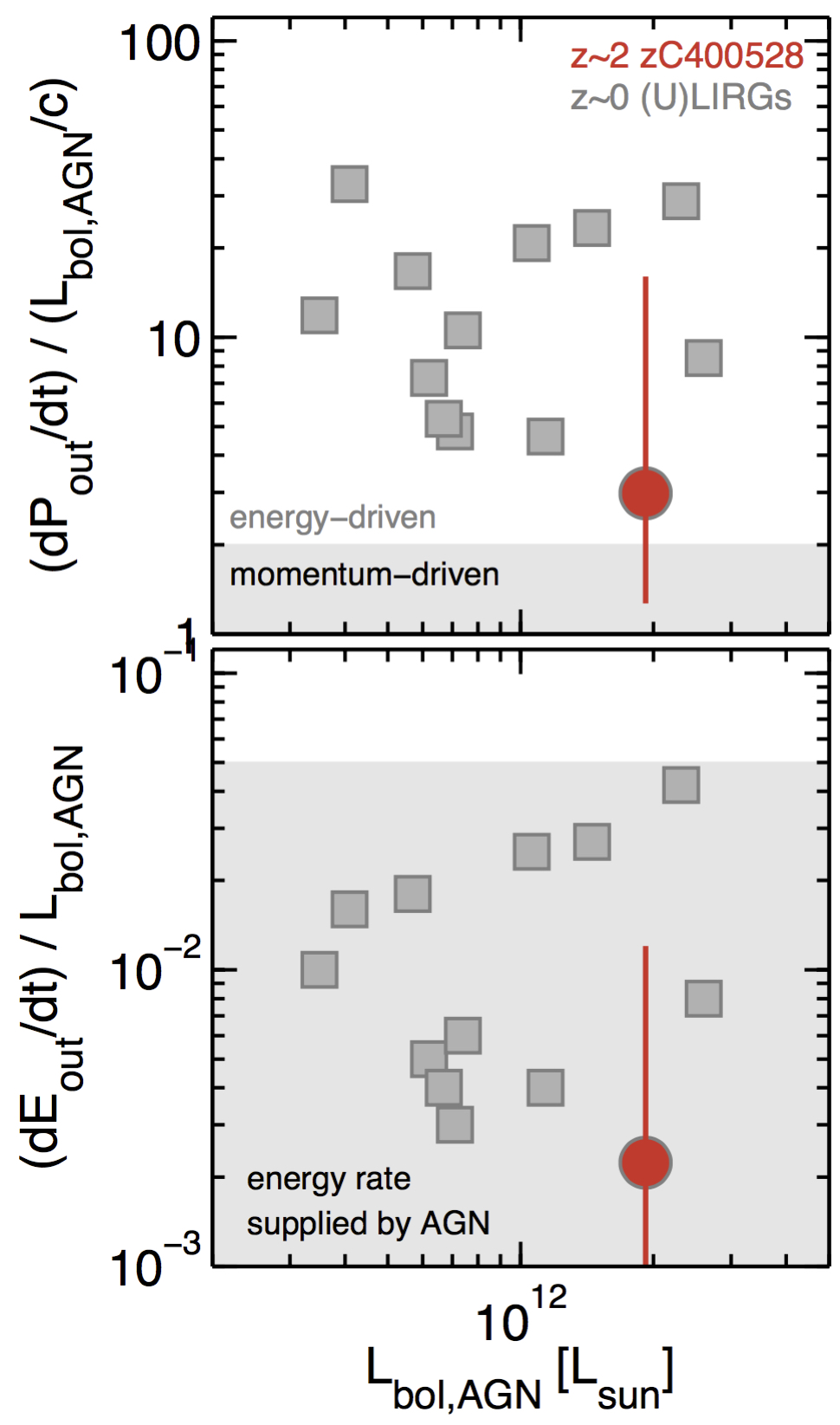}
\caption{Molecular outflow energetics of zC400528 compared to LIRGs from \cite{rhc_gonzalez-alfonso17}. ({\it Top panel}) Momentum flux normalized by the radiation momentum rate of the AGN ($L_{\rm bol,AGN}/c$) as a function of the AGN bolometric luminosity. ({\it Bottom panel}) Mechanical power ($\dot{E}_{\rm out}$) normalized by the AGN bolometric luminosity as a function of the latter. The shaded rectangles mark the momentum and energy rates that can be supplied by an AGN. The vertical red lines associated with the outflow energetics of zC400528 show how the values would change if we assume an optically thin or MW CO-to-H$_{2}$ conversion factor instead of $\alpha_{\rm CO,ULIRG}$.}\label{energetics}
\end{center}
\end{figure}

\begin{figure*}
\begin{center}
\includegraphics[scale=0.24]{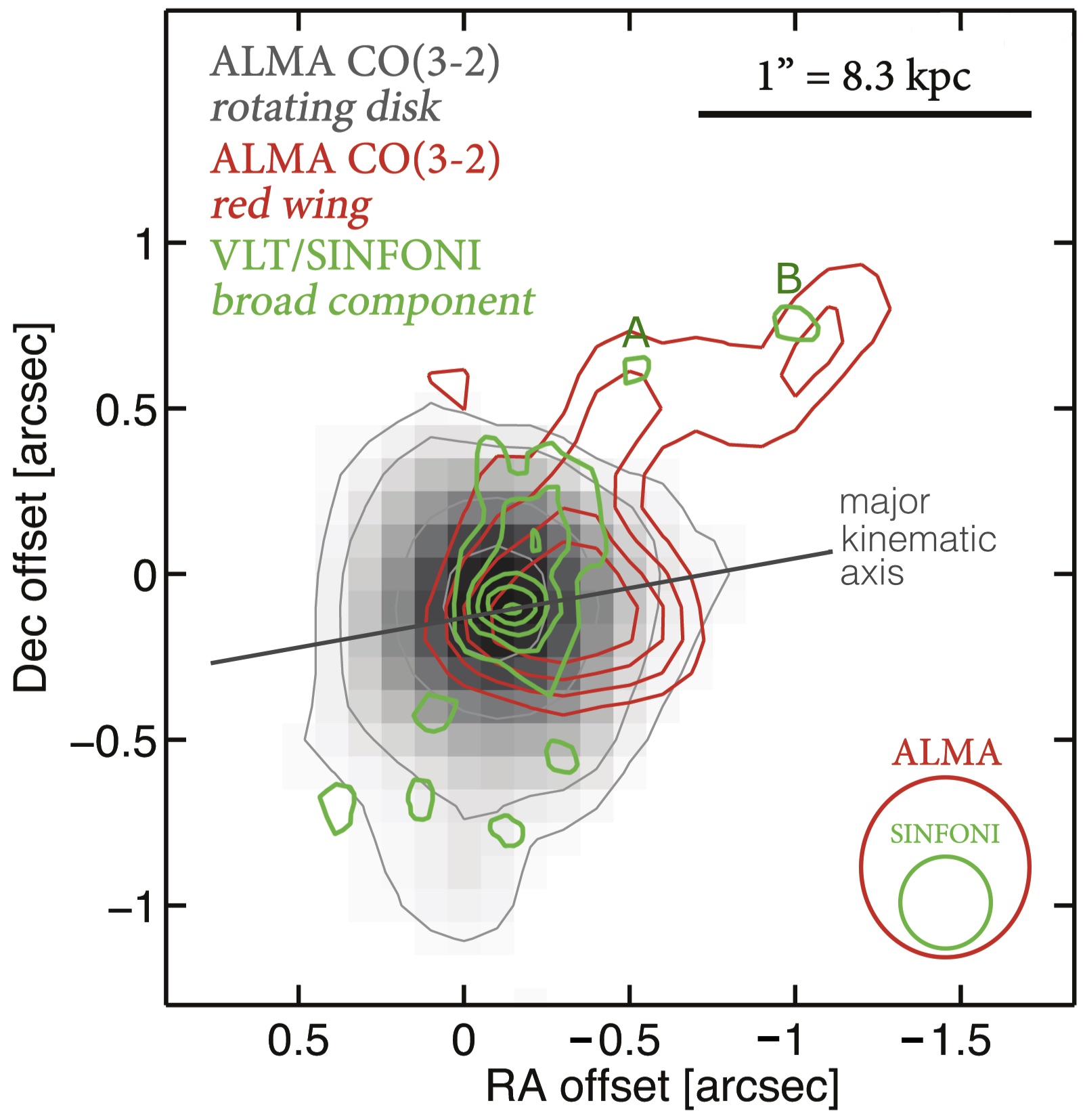} 
\caption{Spatial distribution of the molecular (red) and ionized \citep[green;][]{rhc_f-s14} outflow components of zC400528 overlaid on a map of its molecular disk as traced by the ALMA CO(3-2) data. The grey solid line shows the orientation of the kinematic major axis, the black horizontal bar shows the spatial scale, and the ellipses in the lower-right corner show the angular resolution achieved by the high-resolution SINFONI+AO \citep{rhc_f-s18} and ALMA observations. The letters A and B are assigned to extra-nuclear regions ($\gtrsim4$~kpc) in the outflow that are detected in both molecular and ionized gas. As Figure~\ref{Enuc} in the Appendix shows, the central velocities of the CO and H$\alpha$ lines in these extra-nuclear regions are comparable.}\label{outflow_mom0}
\end{center}
\end{figure*}

\subsection{Energetics of the molecular outflow}\label{section:energetics}

We calculate the momentum flux ($\dot{P}_{\rm out}$) and mechanical power ($\dot{E}_{\rm out}$) in the outflow following $\dot{P}_{\rm out} = \dot{M}_{\rm out} \times v_{\rm out}$ and $\dot{E}_{\rm out} = \frac{1}{2} \dot{M}_{\rm out} \times v_{\rm out}^2$.

From the mass outflow rate $\dot{M}_{\rm out}^{\rm avg}$ calculated in \S\ref{section:Mdot}, the momentum flux is

\begin{equation*}
\begin{split}
\dot{P}_{\rm out,mol}\approx7.2\times10^{35}~{\rm dynes}\times\bigg(\frac{\alpha_{\rm CO,out}}{\alpha_{\rm CO,ULIRG}}\bigg)\\
\times\bigg(\frac{v_{\rm out}}{450~{\rm km}~{\rm s}^{-1}}\bigg)^2\times\bigg(\frac{4.2~{\rm kpc}}{R_{\rm out}}\bigg)
\end{split}
\end{equation*}

\noindent and the mechanical luminosity is

\begin{equation*}
\begin{split}
\dot{E}_{\rm out,mol}\approx1.6\times10^{43}~{\rm erg~s}^{-1}\times\bigg(\frac{\alpha_{\rm CO,out}}{\alpha_{\rm CO,ULIRG}}\bigg)\\
\times\bigg(\frac{v_{\rm out}}{450~{\rm km}~{\rm s}^{-1}}\bigg)^3\times\bigg(\frac{4.2~{\rm kpc}}{R_{\rm out}}\bigg).
\end{split}
\end{equation*}

\noindent Relative to the AGN bolometric luminosity $(L_{\rm AGN,bol}; see Appendix~\ref{agn_bol})$ and radiation momentum rate ($L_{\rm AGN,bol}/{\rm c}$) measured in zC400528, the momentum boost is $\dot{P}_{\rm out,mol}/(L_{\rm AGN,bol}/c)\approx3$ and $\dot{E}_{\rm out,mol}\approx0.2\%L_{\rm AGN,bol}$.

As Figure~\ref{energetics} shows, the momentum boost in zC400528 is at the low end of the distribution of momentum boost observed in local (U)LIRGs \citep[$\dot{P}_{\rm out}\sim3-30~L_{\rm AGN,bol}/c$;][]{rhc_gonzalez-alfonso17}, and suggest that if the gas in the molecular outflow is not optically thin, then the outflow is not momentum conserving ($\lesssim2 L_{\rm AGN,bol}/c$) but rather energy conserving \citep[e.g.,][]{rhc_zubovas12,rhc_f-g12}. In addition, the outflow energy flux in zC400528 is $0.2\%$ of $L_{\rm bol,AGN}$, in agreement with the theoretical expectations that AGN-driven, energy-conserving bubbles should be able to supply a power up to $\sim5\%$ of $L_{\rm AGN,bol}$ \citep[e.g.,][]{rhc_f-g12,rhc_king15}.

A summary of the molecular outflow mass and energetics measured in zC400528 for different assumptions on $\alpha_{\rm CO}$ can be found in Table~\ref{table:energetics}.

\begin{figure}\label{fig:enuc}
\begin{center}
\includegraphics[scale=0.11]{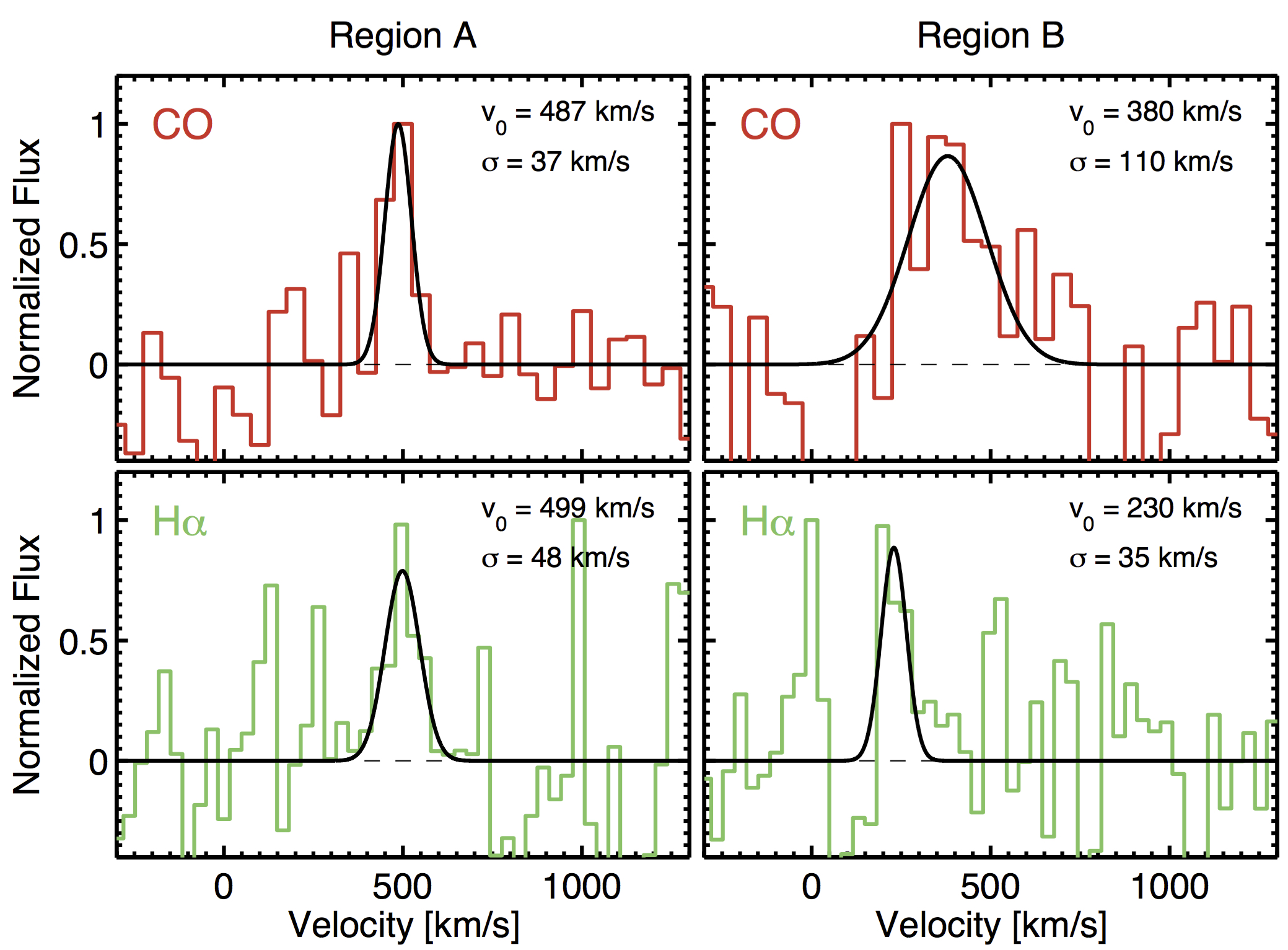}
\caption{CO(3-2) (red) and H$\alpha$ (green) spectra extracted using $\approx0.5$\arcsec apertures centered on extra-nuclear regions A and B (see Figure~\ref{outflow_mom0}) in the outflow of zC400528. The Gaussian fit to the lines is shown in black and the fit central velocity and linewidth ($\sigma$) values are listed in the top-right corner of each panel.} \label{Enuc}
\end{center}
\end{figure}

\begin{figure*}
\begin{center}
\includegraphics[scale=0.085]{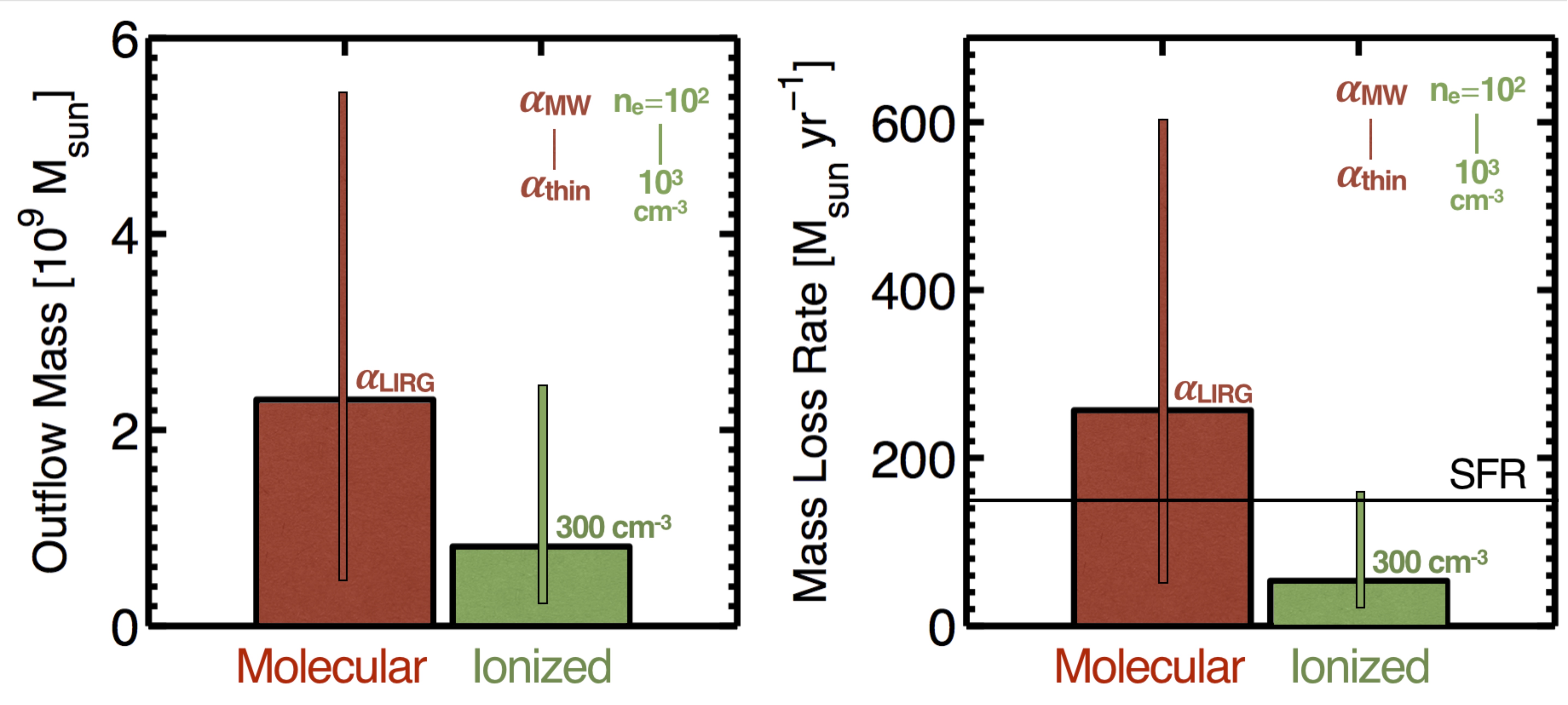}
\caption{{\it (Left)} Outflow mass in the molecular (red) and ionized (green) gas phases calculated assuming a CO-to-H$_{2}$ conversion factor $\alpha_{\rm CO,ULIRG}$ and an electron density $n_{\rm e}=300$~cm$^{-3}$. The vertical thin bars show how would the masses increase or decrease if we change our assumptions of $\alpha_{\rm CO}$ and $n_{\rm e}$. {\it (Right)} Similar to the left panel, but this time showing the outflow mass loss rate. The horizontal black line indicates the SFR measured in zC400528.} \label{bar_fraction}
\end{center}
\end{figure*}

\section{Comparison between molecular and ionized outflow}\label{sec:comp}

Our current knowledge of the multi-phase structure and energetics of galactic outflows at high-$z$ remains greatly limited by the scarcity of observations that target both the ionized and neutral phases of the outflow. The subject of this study, zC400528, represents an exception, with high-resolution observations of the ionized (VLT/SINFONI adaptive optics-assisted) and molecular (ALMA) outflow gas available. In this section we use these complementary datasets to compare the spatial distribution and energetics of the ionized and molecular phases of the outflow.

\subsection{Morphology}

Figure~\ref{outflow_mom0} shows the spatial distribution of the molecular disk (background), the \ha\ blue and red broad (outflow) component (green), and the molecular wind (red). The bulk of the projected \ha\ broad emission extends $\sim4$~kpc north from the nuclear region. Beyond this point, the only two extra-nuclear ionized outflow components detected,  A and B, are co-spatial with the molecular outflow gas and at least in the case of region A, as Figure~\ref{Enuc} shows, the molecular and ionized gas velocities are comparable. 

The spatial distribution of the outflow in zC400528 suggests that the ionized gas is outflowing inclined ($\sim60^{\circ}$) with respect to the major kinematic axis, likely along the path of least resistance outside the galaxy plane. There are two main molecular outflow components in the inner $\sim4$~kpc region: one that extends alongside the ionized wind (from the center to region A), and one that  is aligned with the major kinematic axis, and could therefore be equatorial. This picture resembles the structure of the AGN-driven outflow in NGC~1068  \citep{rhc_tecza01,rhc_cecil02}. There the ionized cone is inclined $\sim40^{\circ}$ with respect to the galaxy disk, and the molecular outflow extends preferentially along the galaxy plane launched as a result of the ionization cone sweeping the molecular gas in the inner disk \citep{rhc_garcia-burillo14}. The fact that the outflow in zC400528 appears to be not perpendicular to the disk is a common feature observed in AGN-driven outflows. Star-formation-driven outflows, on the other hand, tend to be align with the minor kinematic axis of the disk \citep[e.g.,][]{rhc_leroy15,rhc_p-s18}.

The configuration of the ionized and molecular phases of the outflow in zC400528 is also in qualitatively agreement with the modeling work of \cite{rhc_zubovas14b}. They find that AGN-driven outflows that start spherical quickly develop a bipolar morphology that expands faster and further in the polar direction (the direction of least resistance) than in the equatorial direction. After $\sim10$~Myr (the flow timescale in zC400528), several cold dense clumps have formed embedded in the hot gas and are moving upward, while in the plane of the galaxy, the cold gas, squeezed by the expanding bubbles, is being pushed outwards with a mean velocity of $\sim400$~km~s$^{-1}$.

\subsection{Mass and energetics}

The ionized mass in the outflow of zC400528 is $M_{\rm out,ion}=2.2\times(300~{\rm cm}^{-3}/n_{\rm e})\times10^8~M_{\odot}$ \citep{rhc_genzel14}, where $n_{\rm e}$ corresponds to the electron density of the ionized gas in the outflow. Recent results by \cite{rhc_f-s18b} find that the mean electron density in the outflow gas of $z\sim2$ galaxies is $n_{\rm e}\sim350$~cm$^{-3}$.

For an outflow size of $R_{\rm out,ion}=3$~kpc \citep{rhc_f-s14} and an outflow velocity of $v_{\rm out,ion}=802$~km~s$^{-1}$ \citep{rhc_genzel14}, the ionized mass outflow rate is  $\dot{M}_{\rm ion,out}=53$~M$_{\odot}~{\rm yr}^{-1}$. Based on the same assumptions, the ionized momentum and energy outflow rates are $\dot{P}_{\rm out,ion}=2.7\times10^{35}$~dynes and $\dot{E}_{\rm out,ion}=1.1\times10^{43}$~erg~s$^{-1}$, respectively. Figure~\ref{bar_fraction}  summarizes the mass and mass loss rates measured in the molecular and ionized outflow gas in zC400528. The molecular phase dominates the total {\it observed} budget of both $M_{\rm out}$ and $\dot{M}_{\rm out}$. Note, however, that there are gas phases in the outflow that remain unobserved and could change the overall balance, including very hot gas that could significantly contribute to the ionized outflow mass \citep[e.g.,][]{rhc_veilleux05}.

A summary of the ionized outflow mass and energetics measured in zC400528 for different assumptions on $n_{\rm e}$, and the comparison to the quantities measured in the molecular phase, can be found in Table~\ref{table:energetics}.

\begin{figure}
\begin{center}
\includegraphics[scale=0.125]{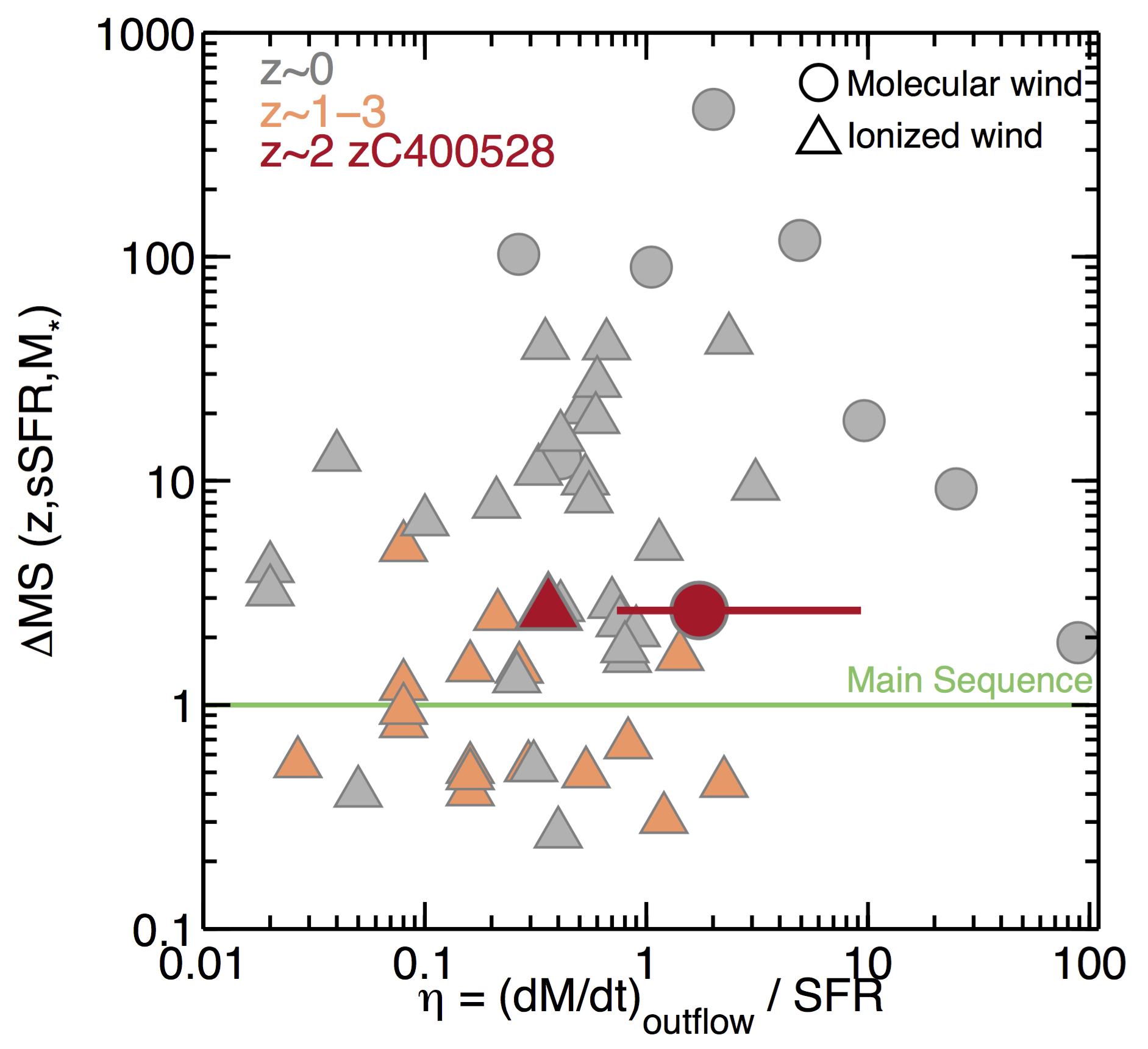}
\caption{Separation from the main-sequence of galaxies (after removing the dependence with $z$ using \cite{rhc_tacconi18} scaling relations) versus their molecular and ionized mass loading factors ($\eta\equiv{\dot M}_{\rm out}/{\rm SFR}$). The circles and triangles show the results for molecular and ionized winds, respectively. In the case of zC400528, the molecular (this work) and ionized \citep{rhc_genzel14,rhc_f-s14} outflow components are shown in red. The vertical red line shows how the molecular gas quantities would change if we assume an optically thin or MW CO-to-H$_{2}$ conversion factor instead of $\alpha_{\rm CO,ULIRG}$. We also include massive, star-forming galaxies in the redshift range $z\sim1-3$ from \cite{rhc_genzel14}, and nearby starburst and AGN galaxies from \cite{rhc_heckman15} and \cite{rhc_fiore17}. }\label{eta}
\end{center}
\end{figure}

Another quantity of interest in the study of galactic outflows is the mass loading factor ($\eta_{\rm phase}$), which is defined as a the ratio between the mass outflow rate and the SFR. Figure~\ref{eta} shows the molecular and ionized mass loading factors as a function of the separation from the main-sequence of galaxies (removing the $z$ dependence using \cite{rhc_tacconi18} scaling relations) for nearby starbursts and AGN galaxies \citep[][]{rhc_heckman15,rhc_fiore17} and massive (log$(M_{*}/M_{\odot})\geq10.9$), star-forming galaxies at $z\sim1-3$ \citep[this work; ][]{rhc_genzel14}. The ionized mass loading factors of these high-$z$ galaxies\footnote{The ionized gas outflow masses reported in \cite{rhc_genzel14} are calculated assuming  $n_{\rm e}=80$~cm$^{-3}$ in the ionized outflowing gas. Here we rescaled those values to match our assumption of $n_{\rm e}=300$~cm$^{-3}$.} are found to be comparable to those observed in local main-sequence outliers.

The molecular mass outflow rate of zC400528 is higher than its SFR ($\eta_{\rm mol}=1.7$), and comparable to some of the powerful molecular outflows observed in local starbursts and Seyfert galaxies (see also \S\ref{section:Mdot}). In addition, the molecular mass loading factor is $\sim4$ times higher than its ionized counterpart. Note, however, that these results depend on assumptions of the properties of the outflow gas that are not yet fully constrained (e.g., the CO-to-H$_2$ conversion factor, the electron density, geometry, etc). The same caveat applies to all the other measurements shown in the figure. 

\begin{deluxetable*}{ccccc}
\tabletypesize{\footnotesize} 
\tablecaption{Molecular and Ionized gas outflow Masses and Energetics\label{table:energetics}}
\tablehead{
\colhead{Phase} & \colhead{Mass} & \colhead{$\dot{M}_{\rm out}$} & \colhead{$\dot{P}_{\rm out}$} & \colhead{$\dot{E}_{\rm out}$} \\
\colhead{} & \colhead{$10^9~M_{\odot}$} & \colhead{$M_{\odot}~{\rm yr}^{-1}$} & $10^{35}$~dynes & $10^{43}$~erg~$s^{-1}$}  
\startdata
Molecular\tablenotemark{a} & 3.36 $(1.42-18.06)$ & 256 $(108.8-1376.0)$ & 7.2 $(3.1-38.7)$ & 1.6 $(0.68-8.6)$ \\
Ionized\tablenotemark{b} & 0.22 $(0.07-0.66)$ & 53 $(15.9-159.0)$ & 2.7 $(0.81-8.1)$ & 1.10 $(0.33-3.30)$ \\ 
\enddata
\tablenotetext{a}{For molecular gas quantities we assume an $\alpha_{\rm CO,ULIRG}$ conversion factor. The values in parenthesis correspond to the values we would obtain if we apply a $\alpha_{\rm CO,thin}$ or $\alpha_{\rm CO,MW}$ conversion factor, respectively.}
\tablenotetext{b}{For ionized gas quantities we assume $n_{\rm e}=300~{\rm cm}^{-3}$ following \cite{rhc_f-s18b}. The values in parenthesis correspond to the values we would obtain if we assume $n_{\rm e}=10^3~{\rm cm}^{-3}$ or $n_{\rm e}=100~{\rm cm}^{-3}$, respectively.}
\end{deluxetable*}

\section{Discussion}\label{sec:escape}

\subsection{Can the outflowing material escape the gravitational potential of its host?}\label{section:escape}

Understanding whether baryons ejected by winds escape the galaxy or rain back to the disk is key to reproduce the observed galaxy stellar masses and the metal enrichment of the IGM \citep[e.g.,][]{rhc_oppenheimer10,rhc_muratov15}. Simulations show that about half of the ejected outflow mass across all galaxy masses is later re-accreted \citep{rhc_uebler14,rhc_christensen16}, and that in massive $z\sim2$ galaxies about $\sim30-40\%$ of the stellar mass forms from gas contributed by wind recycling \citep{rhc_a-c17}.

To determine if the molecular gas in the outflow of zC400528 can permanently escape the galaxy, first we estimate the escape velocity ($v_{\rm esc}$) from the host. For this we follow \cite{rhc_rupke02} and use a simple gravitational model based on a truncated ($r<r_{\rm max}$) isothermal sphere, so $v_{\rm esc}(r) = \sqrt{2}v_{\rm circ}[1+{\rm ln}(r_{\rm max}/r)]^{1/2}$ \footnote{The escape velocity is only weakly sensitive to the exact value of $r_{\rm max}/r$.}. Assuming that the dark matter halo extends to $r_{\rm max}\sim100$~kpc and using a circular velocity for zC400528 of $v_{\rm circ}=344$~km~s$^{-1}$ \citep{rhc_f-s18}, the escape velocity at $r=10$~kpc is $v_{\rm esc, 10kpc}\sim880$~km~s$^{-1}$. The highest velocity molecular outflow material we detect in zC400528 has a velocity of $v_{\rm out}\approx500$~km~s$^{-1}$. If we take into account inclination effects (see \S\ref{section:Mdot}) the deprojected outflow velocity could increase to $v_{out}\sim650$~km~s$^{-1}$, which is still lower than the escape velocity.  This suggests that the bulk of the expelled molecular gas mass will be re-accreted back onto the galaxy on timescales --according to simulations-- no shorter than $\sim1$~Gyr  \citep{rhc_oppenheimer10,rhc_uebler14,rhc_christensen16}. 

In the case of the ionized outflowing gas, the escape velocity at 3~kpc --the distance about the size of the ionized outflow-- is $v_{\rm esc, 3kpc}\sim1,000$~km~s$^{-1}$. The highest velocity gas in the ionized wind can reach velocities up to $\sim800$~km~s$^{-1}$ \citep{rhc_genzel14}, which if we correct by inclination can increase to $\sim1,000$~km~s$^{-1}$ and potentially escape the galaxy.

The very small fraction of the molecular and ionized outflow gas that can escape the gravitational field of zC400528 is consistent with the low escape fractions ($\lesssim20\%$) measured in neutral, ionized and molecular gas winds in local (U)LIRGs \citep[e.g.,][]{rhc_rupke05,rhc_arribas14,rhc_p-s18,rhc_fluetsch18}. 

\subsection{AGN flickering and outflow expansion}\label{section:burst}

Observational and theoretical arguments suggest that the AGN lifetime of $\sim10^{7}-10^{9}$~yr is in reality a succession of hundreds or thousands of short ($\sim10^{5}$~yr) phases of supermassive black hole growth \citep[e.g.,][]{rhc_schawinski15} -- a process called AGN flickering. These periods of episodic AGN activity inflate large-scale outflows which can continue to expand even after the AGN finally switches off as a result of clearing the gas in the central region. 

\cite{rhc_zubovas16} used an analytical model to follow the evolution of the molecular and ionized phases in an outflow driven by an AGN that `flickers' on timescales of $5\times10^{4}$~yr. They find that the oscillations caused by the AGN flickering smooth out relatively fast ($\sim3$~Myr), so from an observational standpoint it would appear that the outflow is driven by an AGN with constant luminosity. For a massive galaxy similar to zC400528 \citep[$M_{\rm h}\approx10^{12.5-13}~M_{\odot}$ based on the relation between halo and stellar mass in ][]{rhc_moster13} the model predicts that in the outflow timescale of zC400528 ($\sim10^7$~yr) the wind should have expanded between $\sim4$ and 8~kpc with expansion velocities at that distance of $\sim300$ and 700~km~s$^{-1}$ for AGN duty cycles of $f_{\rm AGN}=20\%$ and 100\%, respectively. The observed extent ($\sim8$~kpc) and velocity ($\sim500$~km~s$^{-1}$) in the molecular outflow of zC400528 is consistent with these model results and favors the scenario with a high AGN duty cycle, in agreement with the results of \cite{rhc_genzel14} and \cite{rhc_f-s18b}.

\subsection{Is zC400528 moving down from the main-sequence of star-forming galaxies to join the population of compact, passive galaxies?}\label{sec:quenching}

There is growing evidence that suggests that when massive, star-forming galaxies at $z\approx2$ reach central stellar surface densities similar to those observed in massive ellipticals at $z\approx0$ ($\Sigma_{\star}\gtrsim10^{10}$~M$_{\odot}$~kpc$^{-2}$ inside the inner $\sim$kiloparsec region), then an effective quenching mechanism must act on relatively short timescales to shut down the star formation activity \citep[e.g.,][]{rhc_tacchella15b}. AGN-feedback, prevalent in massive $z\sim2$ galaxies \citep[e.g.,][]{rhc_genzel14,rhc_f-s18b}, is one of the obvious quenching candidates.

zC400528 is one of the galaxies in the study of \cite{rhc_tacchella15b} that shows  evidence for inside-out quenching on timescales $\lesssim1$~Gyr in the inner $\sim3$~kpc region. The fact that we detect a powerful nuclear outflow in zC400528 that is removing the molecular gas on a timescale of only half a billion years argues strongly in favor of the outflow as one of the main internal mechanisms responsible for the central suppression of the star formation activity. This conclusion is reinforced if we consider that the nuclear outflow is in reality not capable of clearing the whole galaxy of its star-forming material, but only the central region. If we recalculate the outflow depletion timescale only considering the molecular gas mass in the inner $\sim 4$~kpc region we obtain a depletion timescale that is a factor of $\sim3$ shorter ($t_{\rm dep,out}^{\rm 4~kpc}\approx1.5\times10^{8}$~yr). Note that for the nuclear outflow in zC400528 to be responsible for the quenching of the star-formation activity in the {\it entire} galaxy would require an efficient mode of ``preventive'' feedback, i.e., that the outflow injects enough energy into the halo to keep it hot, drastically reducing the accretion of gas into the disk, including the recycling of molecular gas expelled by the nuclear outflow \citep[e.g.,][]{rhc_gabor11,rhc_king15,rhc_costa17,rhc_dave17}. 

Keeping in mind the many uncertainties associated to the calculation of the mass outflow rate described in \S4, we can attempt to further explore the quenching scenario with a simple back-of-the-envelope calculation. We start by estimating the balance of gas input and drainage $B={\rm inflow~rate/(outflow~rate + SFR)}$, which is one the key parameters regulating the movement of galaxies relative to the main-sequence. To calculate the inflow rate, we use the expression for the mean gas accretion rate in \cite{rhc_bouche10} \citep[see also][]{rhc_lilly13}:

\begin{equation}
\dot{M}_{\rm gas,in}\simeq 90~\epsilon_{\rm in}~f_{\rm b,0.18}~(M_{\rm h,12})^{1.1}~\bigg(\frac{1+z}{3.2}\bigg)^{2.2}~M_{\odot}~{\rm yr}^{-1},
\end{equation}

\noindent where $\epsilon_{\rm in}$ is the accretion efficiency, $f_{\rm b,0.18}$ ($\equiv f_{\rm b}/0.18$) is the cosmic baryonic fraction, and $M_{\rm h,12}$ $(\equiv M_{\rm h}/10^{12}~M_{\odot})$ is the mass of the halo. Assuming a fiducial value for the accretion efficiency of $\epsilon_{\rm in}=0.7$ \citep[e.g.,][]{rhc_bauermesiter10}, a cosmic baryonic fraction of $f_{\rm b}=0.18$, and a halo mass of $M_{\rm h}\approx10^{12.5}~M_{\odot}$ \citep[based on the relation between halo and stellar mass in ][]{rhc_moster13}, we obtain a mean gas accretion rate for zC400528 of $\dot{M}_{\rm gas,in}\approx240~M_{\odot}~{\rm yr}^{-1}$. Then, the balance of gas input and drainage in zC400528 is 

\begin{equation*}
B=\frac{240~M_{\odot}~{\rm yr}^{-1}}{(148+256)~M_{\odot}~{\rm yr}^{-1}}=0.6. 
\end{equation*}

\noindent From the scaling relation described in \cite{rhc_tacchella16} between the balance term $B$ (in the central 5~kpc) and the rate of change of distance from the main-sequence, $\dot{\Delta}_{\rm MS}$, we speculate that zC400528 is {\it moving down from} the main-sequence at a rate of $\dot{\Delta}_{\rm MS}\sim1$~dex~Gyr$^{-1}$. At this rate, and ignoring other mechanisms that could contribute to the star formation activity on timescales $\lesssim1$~Gyr (e.g., re-accretion of ejected molecular gas expelled by the outflow; see \S\ref{section:escape}), it should only take zC400538 about $\sim2$~Gyr to transition from the main-sequence to the regime of passive, quenched galaxies $\sim2$~dex below the main-sequence. This is consistent with the expectations from inside-out quenching that claims that massive, star-forming galaxies such as zC400528 should be fully quenched by $z\sim1$ ($\sim3$~Gyr later than the $z\sim2$ epoch).

\section{Conclusions} \label{sec:conclusions}

In this paper we use ALMA CO(3-2) observations to study the molecular gas properties of the disk and the AGN-driven outflow in zC400528, a {\it typical} massive galaxy at $z=2.3$ in the process of quenching \citep{rhc_tacchella15}. We complement these observations with VLT/SINFONI adaptive-optics \ha\ data \citep[][]{rhc_f-s14,rhc_f-s18} which allowed us to conduct one of the first spatially-resolved, multi-phase studies of a galactic outflow at $z\sim2$.

We highlight the following points: 

\begin{enumerate}
\item {{\bf Molecular disk:} The molecular gas mass in the rotating disk of zC400528 (assuming a CO-to-H$_2$ conversion factor of $\alpha_{\rm CO(1-0),T18}=4.3~M_{\odot}~({\rm K~km~s^{-1}~pc^{-2}})^{-1}$) is $M_{\rm mol,disk}=1.1\times10^{11}$~$M_{\odot}$, which translates to a molecular gas to stellar ratio of $M_{\rm mol}/M_{\star}\approx1$. This high gas mass fraction is comparable to those observed in massive star-forming galaxies at $z\sim2$ \citep[][]{rhc_genzel15,rhc_tacconi18}.
}

\item {{\bf Molecular outflow:} The molecular outflow in zC400528 is asymmetrical, with only its receding component detected out to a projected distance of $\sim10$~kpc. The bulk of the emission is concentrated in the nuclear region and aligned with the major kinematic axis of the disk. We observe a positive velocity gradient in the outflow with projected velocities that range from $\sim300$~km~s$^{-1}$ in the nuclear part to $\sim400-500$~km~s$^{-1}$ in the furthest components. 

We measure a molecular outflow mass and mass loss rate of $M_{\rm out,mol}=3.36\times10^{9}~M_{\odot}$ and $\dot{M}_{\rm out,mol}^{\rm avg}\approx256~M_{\odot}~{\rm yr^{-1}}$, respectively (these assume a ULIRG-like CO-to-H$_2$ conversion factor). With this powerful molecular mass loss rate zC400528 is depleting its nuclear molecular gas twice as fast as due to its star formation activity. 

The momentum boost in the molecular phase of the outflow is $\dot{P}_{\rm out,mol}/(L_{\rm AGN,bol}/c)\approx3$ and the energy flux is $\dot{E}_{\rm out,mol}\approx0.2\%L_{\rm AGN,bol}$, in agreement with theoretical expectations for an AGN-driven, energy-conserving outflow.}

\item{{\bf Molecular and ionized phases in the outflow:} 
The ionized and molecular phases of the outflow in zC400528 have different morphological properties. While the ionized outflow extends for $\sim4$~kpc from the nuclear region inclined about $60^{\circ}$ with respect to the major kinematic axis of the disk, the central component of the molecular outflow is found preferentially in between the ionized wind and the major kinematic axis. 
This type of configuration is not uncommon in AGN-driven outflows as found in observational and simulation studies \citep[e.g.,][]{rhc_garcia-burillo14,rhc_zubovas14a}. We also observe that in the extended molecular component of the outflow ($\gtrsim5$~kpc) there are at least two regions (A and B in Fig.~\ref{outflow_mom0}) where ionized and molecular gas are co-spatial and moving at similar high-velocities.

Keeping in mind all the uncertainties that affect the determination of the ionized and molecular outflow properties (e.g., electron density, CO-to-H$_2$ conversion factor, geometry, etc.), both the mass and energetics of the outflow are dominated by the molecular phase. For the ionized phase to be dominant would require the combination of low electron densities and optically thin molecular gas in the  wind. Both of these conditions seem unlikely \citep[e.g., ][]{rhc_aalto15,rhc_walter17,rhc_f-s18b}. The mass loading factors ($\dot{M}_{\rm out}/{\rm SFR}$) of the molecular and ionized outflow phases are 1.7 and 0.4, respectively.}

\item{{\bf AGN-driven outflow quenching in action?} Kilo-parsec scale observations of the stellar content and the star formation activity in zC400528 reveal that one or more quenching mechanisms are shutting down the star formation activity from the inside-out on short timescales \citep[$\lesssim1.5$~Gyr; ][]{rhc_tacchella15b}. The fact that we detect an AGN-driven outflow powerful enough to expel the star-forming material in the central region on short timescales ($t_{\rm dep,out}\sim0.2$~Gyr if we consider the molecular gas in the inner four kiloparsecs) points at the outflow as one of the main quenching engines at work. Although it is true that the expelled molecular gas is not fast enough to escape the system, the energy input of the outflow into the halo should reduce the accretion of both fresh and recycled gas into the disk, facilitating even more the transition of zC400528 into the realm of passive systems below the main-sequence of galaxies.}

\end{enumerate}

\section{Acknowledgments}

RHC would like to thank the support and encouragement from Fares Bravo Garrido and dedicates this paper with love to Fares and Olivia. SGB thanks economic support from funding grant AYA2016-76682-C3-2-P. CF acknowledges support from the European Union Horizon 2020 research and innovation programme under the Marie Sklodowska-Curie grant agreement No 664931. LC acknowledges financial support by the Spanish Ministry for Science, Innovation and Universities under grants AYA2015-68964 and ESP2017-83197. The Cosmic Dawn Center is funded by the Danish National Research Foundation. This paper makes use of the following ALMA data: ADS/JAO.ALMA\#2013.1.00092.S and \#2015.1.00220.S. ALMA is a partnership of ESO (representing its member states), NSF (USA) and NINS (Japan), together with NRC (Canada) and NSC and ASIAA (Taiwan) and KASI (Republic of Korea), in cooperation with the Republic of Chile. The Joint ALMA Observatory is operated by ESO, AUI/NRAO and NAOJ. This work is based on observations taken by the 3D-HST Treasury Program (GO 12177 and 12328) with the NASA/ESA HST, which is operated by the Association of Universities for Research in Astronomy, Inc., under NASA contract NAS5-26555.

\appendix

\section{AGN bolometric luminosity of zC400528}\label{agn_bol}

Given the non-detection of zC400528 in X-ray emission described in \S\ref{galaxy_info}, we estimate its AGN bolometric luminosity in two ways: (1) We convert its SFR into a total infrared luminosity ($L_{\rm TIR}$) using the prescription by \cite{rhc_murphy11}. Then, we use the common assumption that $L_{\rm bol}=1.15 L_{\rm TIR}$, and that a fraction $f_{\rm AGN}=0.5$ \citep[a typical AGN fraction in ULIRGs, e.g.,][]{rhc_veilleux09} of $L_{\rm bol}$ is associated to the bolometric AGN luminosity. This results in $log_{10}(L_{\rm AGN, bol}/[{\rm erg~s}^{-1}])\approx45.3$. (2) We follow the methodology described in \cite{rhc_f-s18b} and use the \nii$\lambda$6584 luminosity in the narrow component after subtracting the contribution by star-forming regions based on the mass-metallicity relation. The resulting AGN bolometric luminosity is $log_{10}(L_{\rm AGN, bol}/[{\rm erg~s}^{-1}])\approx46.1$, a factor of $\sim6$ higher than the TIR-based estimate in (1). In the analysis we use the average between these two values, i.e., $log_{10}(L_{\rm AGN, bol}/[{\rm erg~s}^{-1}])\approx45.8$.

\bibliography{references.bib}

\begin{thebibliography}{}
\expandafter\ifx\csname natexlab\endcsname\relax\def\natexlab#1{#1}\fi

\bibitem[{{Aalto} {et~al.}(2012){Aalto}, {Garcia-Burillo}, {Muller}, {Winters},
  {van der Werf}, {Henkel}, {Costagliola}, \& {Neri}}]{rhc_aalto12}
{Aalto}, S., {Garcia-Burillo}, S., {Muller}, S., {et~al.} 2012, \aap, 537, A44

\bibitem[{{Aalto} {et~al.}(2015){Aalto}, {Garcia-Burillo}, {Muller}, {Winters},
  {Gonzalez-Alfonso}, {van der Werf}, {Henkel}, {Costagliola}, \&
  {Neri}}]{rhc_aalto15}
---. 2015, \aap, 574, A85

\bibitem[{{Alatalo}(2015)}]{rhc_alatalo15}
{Alatalo}, K. 2015, \apjl, 801, L17

\bibitem[{{Angl{\'e}s-Alc{\'a}zar} {et~al.}(2017){Angl{\'e}s-Alc{\'a}zar},
  {Faucher-Gigu{\`e}re}, {Kere{\v s}}, {Hopkins}, {Quataert}, \&
  {Murray}}]{rhc_a-c17}
{Angl{\'e}s-Alc{\'a}zar}, D., {Faucher-Gigu{\`e}re}, C.-A., {Kere{\v s}}, D.,
  {et~al.} 2017, \mnras, 470, 4698

\bibitem[{{Arribas} {et~al.}(2014){Arribas}, {Colina}, {Bellocchi}, {Maiolino},
  \& {Villar-Mart{\'{\i}}n}}]{rhc_arribas14}
{Arribas}, S., {Colina}, L., {Bellocchi}, E., {Maiolino}, R., \&
  {Villar-Mart{\'{\i}}n}, M. 2014, \aap, 568, A14

\bibitem[{{Barcos-Mu{\~n}oz} {et~al.}(2018){Barcos-Mu{\~n}oz}, {Aalto},
  {Thompson}, {Sakamoto}, {Mart{\'{\i}}n}, {Leroy}, {Privon}, {Evans}, \&
  {Kepley}}]{rhc_barcos18}
{Barcos-Mu{\~n}oz}, L., {Aalto}, S., {Thompson}, T.~A., {et~al.} 2018, \apjl,
  853, L28

\bibitem[{{Barro} {et~al.}(2016){Barro}, {Kriek}, {P{\'e}rez-Gonz{\'a}lez},
  {Trump}, {Koo}, {Faber}, {Dekel}, {Primack}, {Guo}, {Kocevski},
  {Mu{\~n}oz-Mateos}, {Rujopakarn}, \& {Seth}}]{rhc_barro16}
{Barro}, G., {Kriek}, M., {P{\'e}rez-Gonz{\'a}lez}, P.~G., {et~al.} 2016,
  \apjl, 827, L32

\bibitem[{{Bauermeister} {et~al.}(2010){Bauermeister}, {Blitz}, \&
  {Ma}}]{rhc_bauermesiter10}
{Bauermeister}, A., {Blitz}, L., \& {Ma}, C.-P. 2010, \apj, 717, 323

\bibitem[{{Bolatto} {et~al.}(2013{\natexlab{a}}){Bolatto}, {Wolfire}, \&
  {Leroy}}]{rhc_bolatto13}
{Bolatto}, A.~D., {Wolfire}, M., \& {Leroy}, A.~K. 2013{\natexlab{a}}, \araa,
  51, 207

\bibitem[{{Bolatto} {et~al.}(2013{\natexlab{b}}){Bolatto}, {Warren}, {Leroy},
  {Walter}, {Veilleux}, {Ostriker}, {Ott}, {Zwaan}, {Fisher}, {Weiss},
  {Rosolowsky}, \& {Hodge}}]{rhc_bolatto13b}
{Bolatto}, A.~D., {Warren}, S.~R., {Leroy}, A.~K., {et~al.} 2013{\natexlab{b}},
  \nat, 499, 450

\bibitem[{{Bolatto} {et~al.}(2015){Bolatto}, {Warren}, {Leroy}, {Tacconi},
  {Bouch{\'e}}, {F{\"o}rster Schreiber}, {Genzel}, {Cooper}, {Fisher},
  {Combes}, {Garc{\'{\i}}a-Burillo}, {Burkert}, {Bournaud}, {Weiss},
  {Saintonge}, {Wuyts}, \& {Sternberg}}]{rhc_bolatto15}
---. 2015, \apj, 809, 175

\bibitem[{{Bouch{\'e}} {et~al.}(2010){Bouch{\'e}}, {Dekel}, {Genzel}, {Genel},
  {Cresci}, {F{\"o}rster Schreiber}, {Shapiro}, {Davies}, \&
  {Tacconi}}]{rhc_bouche10}
{Bouch{\'e}}, N., {Dekel}, A., {Genzel}, R., {et~al.} 2010, \apj, 718, 1001

\bibitem[{{Cecil} {et~al.}(2002){Cecil}, {Dopita}, {Groves}, {Wilson},
  {Ferruit}, {P{\'e}contal}, \& {Binette}}]{rhc_cecil02}
{Cecil}, G., {Dopita}, M.~A., {Groves}, B., {et~al.} 2002, \apj, 568, 627

\bibitem[{{Christensen} {et~al.}(2016){Christensen}, {Dav{\'e}}, {Governato},
  {Pontzen}, {Brooks}, {Munshi}, {Quinn}, \& {Wadsley}}]{rhc_christensen16}
{Christensen}, C.~R., {Dav{\'e}}, R., {Governato}, F., {et~al.} 2016, \apj,
  824, 57

\bibitem[{{Cicone} {et~al.}(2012){Cicone}, {Feruglio}, {Maiolino}, {Fiore},
  {Piconcelli}, {Menci}, {Aussel}, \& {Sturm}}]{rhc_cicone12}
{Cicone}, C., {Feruglio}, C., {Maiolino}, R., {et~al.} 2012, \aap, 543, A99

\bibitem[{{Cicone} {et~al.}(2014){Cicone}, {Maiolino}, {Sturm},
  {Graci{\'a}-Carpio}, {Feruglio}, {Neri}, {Aalto}, {Davies}, {Fiore},
  {Fischer}, {Garc{\'{\i}}a-Burillo}, {Gonz{\'a}lez-Alfonso},
  {Hailey-Dunsheath}, {Piconcelli}, \& {Veilleux}}]{rhc_cicone14}
{Cicone}, C., {Maiolino}, R., {Sturm}, E., {et~al.} 2014, \aap, 562, A21

\bibitem[{{Concas} {et~al.}(2017){Concas}, {Popesso}, {Brusa}, {Mainieri},
  {Erfanianfar}, \& {Morselli}}]{rhc_concas17}
{Concas}, A., {Popesso}, P., {Brusa}, M., {et~al.} 2017, \aap, 606, A36

\bibitem[{{Contursi} {et~al.}(2013){Contursi}, {Poglitsch}, {Gr{\'a}cia
  Carpio}, {Veilleux}, {Sturm}, {Fischer}, {Verma}, {Hailey-Dunsheath}, {Lutz},
  {Davies}, {Gonz{\'a}lez-Alfonso}, {Sternberg}, {Genzel}, \&
  {Tacconi}}]{rhc_contursi13}
{Contursi}, A., {Poglitsch}, A., {Gr{\'a}cia Carpio}, J., {et~al.} 2013, \aap,
  549, A118

\bibitem[{{Costa} {et~al.}(2017){Costa}, {Rosdahl}, {Sijacki}, \&
  {Haehnelt}}]{rhc_costa17}
{Costa}, T., {Rosdahl}, J., {Sijacki}, D., \& {Haehnelt}, M.~G. 2017, ArXiv
  e-prints, arXiv:1709.08638

\bibitem[{{Daddi} {et~al.}(2015){Daddi}, {Dannerbauer}, {Liu}, {Aravena},
  {Bournaud}, {Walter}, {Riechers}, {Magdis}, {Sargent}, {B{\'e}thermin},
  {Carilli}, {Cibinel}, {Dickinson}, {Elbaz}, {Gao}, {Gobat}, {Hodge}, \&
  {Krips}}]{rhc_daddi15}
{Daddi}, E., {Dannerbauer}, H., {Liu}, D., {et~al.} 2015, \aap, 577, A46

\bibitem[{{Dannerbauer} {et~al.}(2009){Dannerbauer}, {Daddi}, {Riechers},
  {Walter}, {Carilli}, {Dickinson}, {Elbaz}, \& {Morrison}}]{rhc_dannerbauer09}
{Dannerbauer}, H., {Daddi}, E., {Riechers}, D.~A., {et~al.} 2009, \apjl, 698,
  L178

\bibitem[{{Dasyra} {et~al.}(2016){Dasyra}, {Combes}, {Oosterloo}, {Oonk},
  {Morganti}, {Salom{\'e}}, \& {Vlahakis}}]{rhc_dasyra16}
{Dasyra}, K.~M., {Combes}, F., {Oosterloo}, T., {et~al.} 2016, \aap, 595, L7

\bibitem[{{Dav{\'e}} {et~al.}(2017){Dav{\'e}}, {Rafieferantsoa}, \&
  {Thompson}}]{rhc_dave17}
{Dav{\'e}}, R., {Rafieferantsoa}, M.~H., \& {Thompson}, R.~J. 2017, \mnras,
  471, 1671

\bibitem[{{Di Matteo} {et~al.}(2005){Di Matteo}, {Springel}, \&
  {Hernquist}}]{rhc_dimatteo05}
{Di Matteo}, T., {Springel}, V., \& {Hernquist}, L. 2005, \nat, 433, 604

\bibitem[{{Eisenhauer} {et~al.}(2003){Eisenhauer}, {Abuter}, {Bickert},
  {Biancat-Marchet}, {Bonnet}, {Brynnel}, {Conzelmann}, {Delabre}, {Donaldson},
  {Farinato}, {Fedrigo}, {Genzel}, {Hubin}, {Iserlohe}, {Kasper},
  {Kissler-Patig}, {Monnet}, {Roehrle}, {Schreiber}, {Stroebele}, {Tecza},
  {Thatte}, \& {Weisz}}]{rhc_eisenhauer03}
{Eisenhauer}, F., {Abuter}, R., {Bickert}, K., {et~al.} 2003, in \procspie,
  Vol. 4841, Instrument Design and Performance for Optical/Infrared
  Ground-based Telescopes, ed. M.~{Iye} \& A.~F.~M. {Moorwood}, 1548--1561

\bibitem[{{Elvis} {et~al.}(2009){Elvis}, {Civano}, {Vignali}, {Puccetti},
  {Fiore}, {Cappelluti}, {Aldcroft}, {Fruscione}, {Zamorani}, {Comastri},
  {Brusa}, {Gilli}, {Miyaji}, {Damiani}, {Koekemoer}, {Finoguenov}, {Brunner},
  {Urry}, {Silverman}, {Mainieri}, {Hasinger}, {Griffiths}, {Carollo}, {Hao},
  {Guzzo}, {Blain}, {Calzetti}, {Carilli}, {Capak}, {Ettori}, {Fabbiano},
  {Impey}, {Lilly}, {Mobasher}, {Rich}, {Salvato}, {Sanders}, {Schinnerer},
  {Scoville}, {Shopbell}, {Taylor}, {Taniguchi}, \& {Volonteri}}]{rhc_elvis09}
{Elvis}, M., {Civano}, F., {Vignali}, C., {et~al.} 2009, \apjs, 184, 158

\bibitem[{{Faucher-Gigu{\`e}re} \& {Quataert}(2012)}]{rhc_f-g12}
{Faucher-Gigu{\`e}re}, C.-A., \& {Quataert}, E. 2012, \mnras, 425, 605

\bibitem[{{Feruglio} {et~al.}(2010){Feruglio}, {Maiolino}, {Piconcelli},
  {Menci}, {Aussel}, {Lamastra}, \& {Fiore}}]{rhc_feruglio10}
{Feruglio}, C., {Maiolino}, R., {Piconcelli}, E., {et~al.} 2010, \aap, 518,
  L155

\bibitem[{{Feruglio} {et~al.}(2015){Feruglio}, {Fiore}, {Carniani},
  {Piconcelli}, {Zappacosta}, {Bongiorno}, {Cicone}, {Maiolino}, {Marconi},
  {Menci}, {Puccetti}, \& {Veilleux}}]{rhc_feruglio15}
{Feruglio}, C., {Fiore}, F., {Carniani}, S., {et~al.} 2015, \aap, 583, A99

\bibitem[{{Fiore} {et~al.}(2017){Fiore}, {Feruglio}, {Shankar}, {Bischetti},
  {Bongiorno}, {Brusa}, {Carniani}, {Cicone}, {Duras}, {Lamastra}, {Mainieri},
  {Marconi}, {Menci}, {Maiolino}, {Piconcelli}, {Vietri}, \&
  {Zappacosta}}]{rhc_fiore17}
{Fiore}, F., {Feruglio}, C., {Shankar}, F., {et~al.} 2017, \aap, 601, A143

\bibitem[{{Fluetsch} {et~al.}(2018){Fluetsch}, {Maiolino}, {Carniani},
  {Marconi}, {Cicone}, {Bourne}, {Costa}, {Fabian}, {Ishibashi}, \&
  {Venturi}}]{rhc_fluetsch18}
{Fluetsch}, A., {Maiolino}, R., {Carniani}, S., {et~al.} 2018, ArXiv e-prints,
  arXiv:1805.05352

\bibitem[{{F{\"o}rster Schreiber} {et~al.}(2014){F{\"o}rster Schreiber},
  {Genzel}, {Newman}, {Kurk}, {Lutz}, {Tacconi}, {Wuyts}, {Bandara}, {Burkert},
  {Buschkamp}, {Carollo}, {Cresci}, {Daddi}, {Davies}, {Eisenhauer}, {Hicks},
  {Lang}, {Lilly}, {Mainieri}, {Mancini}, {Naab}, {Peng}, {Renzini}, {Rosario},
  {Shapiro Griffin}, {Shapley}, {Sternberg}, {Tacchella}, {Vergani},
  {Wisnioski}, {Wuyts}, \& {Zamorani}}]{rhc_f-s14}
{F{\"o}rster Schreiber}, N.~M., {Genzel}, R., {Newman}, S.~F., {et~al.} 2014,
  \apj, 787, 38

\bibitem[{{F{\"o}rster Schreiber} {et~al.}(2018{\natexlab{a}}){F{\"o}rster
  Schreiber}, {{\"U}bler}, {Davies}, {Genzel}, {Wisnioski}, {Belli}, {Shimizu},
  {Lutz}, {Fossati}, {Herrera-Camus}, {Mendel}, {Tacconi}, {Wilman},
  {Beifiori}, {Brammer}, {Burkert}, {Carollo}, {Davies}, {Eisenhauer},
  {Fabricius}, {Lilly}, {Momcheva}, {Naab}, {Nelson}, {Price}, {Renzini},
  {Saglia}, {Sternberg}, {van Dokkum}, \& {Wuyts}}]{rhc_f-s18b}
{F{\"o}rster Schreiber}, N.~M., {{\"U}bler}, H., {Davies}, R.~L., {et~al.}
  2018{\natexlab{a}}, ArXiv e-prints, arXiv:1807.04738

\bibitem[{{F{\"o}rster Schreiber} {et~al.}(2018{\natexlab{b}}){F{\"o}rster
  Schreiber}, {Renzini}, {Mancini}, {Genzel}, {Bouch{\'e}}, {Cresci}, {Hicks},
  {Lilly}, {Peng}, {Burkert}, {Carollo}, {Cimatti}, {Daddi}, {Davies}, {Genel},
  {Kurk}, {Lang}, {Lutz}, {Mainieri}, {McCracken}, {Mignoli}, {Naab}, {Oesch},
  {Pozzetti}, {Scodeggio}, {Shapiro Griffin}, {Shapley}, {Sternberg},
  {Tacchella}, {Tacconi}, {Wuyts}, \& {Zamorani}}]{rhc_f-s18}
{F{\"o}rster Schreiber}, N.~M., {Renzini}, A., {Mancini}, C., {et~al.}
  2018{\natexlab{b}}, ArXiv e-prints, arXiv:1802.07276

\bibitem[{{Gabor} \& {Bournaud}(2014)}]{rhc_gabor14}
{Gabor}, J.~M., \& {Bournaud}, F. 2014, \mnras, 441, 1615

\bibitem[{{Gabor} {et~al.}(2011){Gabor}, {Dav{\'e}}, {Oppenheimer}, \&
  {Finlator}}]{rhc_gabor11}
{Gabor}, J.~M., {Dav{\'e}}, R., {Oppenheimer}, B.~D., \& {Finlator}, K. 2011,
  \mnras, 417, 2676

\bibitem[{{Garc{\'{\i}}a-Burillo} {et~al.}(2014){Garc{\'{\i}}a-Burillo},
  {Combes}, {Usero}, {Aalto}, {Krips}, {Viti}, {Alonso-Herrero}, {Hunt},
  {Schinnerer}, {Baker}, {Boone}, {Casasola}, {Colina}, {Costagliola},
  {Eckart}, {Fuente}, {Henkel}, {Labiano}, {Mart{\'{\i}}n}, {M{\'a}rquez},
  {Muller}, {Planesas}, {Ramos Almeida}, {Spaans}, {Tacconi}, \& {van der
  Werf}}]{rhc_garcia-burillo14}
{Garc{\'{\i}}a-Burillo}, S., {Combes}, F., {Usero}, A., {et~al.} 2014, \aap,
  567, A125

\bibitem[{{Genzel} {et~al.}(2014){Genzel}, {F{\"o}rster Schreiber}, {Rosario},
  {Lang}, {Lutz}, {Wisnioski}, {Wuyts}, {Wuyts}, {Bandara}, {Bender}, {Berta},
  {Kurk}, {Mendel}, {Tacconi}, {Wilman}, {Beifiori}, {Brammer}, {Burkert},
  {Buschkamp}, {Chan}, {Carollo}, {Davies}, {Eisenhauer}, {Fabricius},
  {Fossati}, {Kriek}, {Kulkarni}, {Lilly}, {Mancini}, {Momcheva}, {Naab},
  {Nelson}, {Renzini}, {Saglia}, {Sharples}, {Sternberg}, {Tacchella}, \& {van
  Dokkum}}]{rhc_genzel14}
{Genzel}, R., {F{\"o}rster Schreiber}, N.~M., {Rosario}, D., {et~al.} 2014,
  \apj, 796, 7

\bibitem[{{Genzel} {et~al.}(2015){Genzel}, {Tacconi}, {Lutz}, {Saintonge},
  {Berta}, {Magnelli}, {Combes}, {Garc{\'{\i}}a-Burillo}, {Neri}, {Bolatto},
  {Contini}, {Lilly}, {Boissier}, {Boone}, {Bouch{\'e}}, {Bournaud}, {Burkert},
  {Carollo}, {Colina}, {Cooper}, {Cox}, {Feruglio}, {F{\"o}rster Schreiber},
  {Freundlich}, {Gracia-Carpio}, {Juneau}, {Kovac}, {Lippa}, {Naab}, {Salome},
  {Renzini}, {Sternberg}, {Walter}, {Weiner}, {Weiss}, \&
  {Wuyts}}]{rhc_genzel15}
{Genzel}, R., {Tacconi}, L.~J., {Lutz}, D., {et~al.} 2015, \apj, 800, 20

\bibitem[{{Gonz{\'a}lez-Alfonso} {et~al.}(2014){Gonz{\'a}lez-Alfonso},
  {Fischer}, {Graci{\'a}-Carpio}, {Falstad}, {Sturm}, {Mel{\'e}ndez}, {Spoon},
  {Verma}, {Davies}, {Lutz}, {Aalto}, {Polisensky}, {Poglitsch}, {Veilleux}, \&
  {Contursi}}]{rhc_gonzalez-alfonso14}
{Gonz{\'a}lez-Alfonso}, E., {Fischer}, J., {Graci{\'a}-Carpio}, J., {et~al.}
  2014, \aap, 561, A27

\bibitem[{{Gonz{\'a}lez-Alfonso} {et~al.}(2017){Gonz{\'a}lez-Alfonso},
  {Fischer}, {Spoon}, {Stewart}, {Ashby}, {Veilleux}, {Smith}, {Sturm},
  {Farrah}, {Falstad}, {Mel{\'e}ndez}, {Graci{\'a}-Carpio}, {Janssen}, \&
  {Lebouteiller}}]{rhc_gonzalez-alfonso17}
{Gonz{\'a}lez-Alfonso}, E., {Fischer}, J., {Spoon}, H.~W.~W., {et~al.} 2017,
  \apj, 836, 11

\bibitem[{{Harrison} {et~al.}(2016){Harrison}, {Alexander}, {Mullaney},
  {Stott}, {Swinbank}, {Arumugam}, {Bauer}, {Bower}, {Bunker}, \&
  {Sharples}}]{rhc_harrison16}
{Harrison}, C.~M., {Alexander}, D.~M., {Mullaney}, J.~R., {et~al.} 2016,
  \mnras, 456, 1195

\bibitem[{{Heckman} {et~al.}(2015){Heckman}, {Alexandroff}, {Borthakur},
  {Overzier}, \& {Leitherer}}]{rhc_heckman15}
{Heckman}, T.~M., {Alexandroff}, R.~M., {Borthakur}, S., {Overzier}, R., \&
  {Leitherer}, C. 2015, \apj, 809, 147

\bibitem[{{Janssen} {et~al.}(2016){Janssen}, {Christopher}, {Sturm},
  {Veilleux}, {Contursi}, {Gonz{\'a}lez-Alfonso}, {Fischer}, {Davies}, {Verma},
  {Graci{\'a}-Carpio}, {Genzel}, {Lutz}, {Sternberg}, {Tacconi}, {Burtscher},
  \& {Poglitsch}}]{rhc_janssen16}
{Janssen}, A.~W., {Christopher}, N., {Sturm}, E., {et~al.} 2016, \apj, 822, 43

\bibitem[{{Kewley} {et~al.}(2013){Kewley}, {Maier}, {Yabe}, {Ohta}, {Akiyama},
  {Dopita}, \& {Yuan}}]{rhc_kewley13}
{Kewley}, L.~J., {Maier}, C., {Yabe}, K., {et~al.} 2013, \apjl, 774, L10

\bibitem[{{King} \& {Pounds}(2015)}]{rhc_king15}
{King}, A., \& {Pounds}, K. 2015, \araa, 53, 115

\bibitem[{{Kornei} {et~al.}(2012){Kornei}, {Shapley}, {Martin}, {Coil}, {Lotz},
  {Schiminovich}, {Bundy}, \& {Noeske}}]{rhc_kornei12}
{Kornei}, K.~A., {Shapley}, A.~E., {Martin}, C.~L., {et~al.} 2012, \apj, 758,
  135

\bibitem[{{Leroy} {et~al.}(2015){Leroy}, {Walter}, {Martini}, {Roussel},
  {Sandstrom}, {Ott}, {Weiss}, {Bolatto}, {Schuster}, \&
  {Dessauges-Zavadsky}}]{rhc_leroy15}
{Leroy}, A.~K., {Walter}, F., {Martini}, P., {et~al.} 2015, \apj, 814, 83

\bibitem[{{Leung} {et~al.}(2017){Leung}, {Coil}, {Azadi}, {Aird}, {Shapley},
  {Kriek}, {Mobasher}, {Reddy}, {Siana}, {Freeman}, {Price}, {Sanders}, \&
  {Shivaei}}]{rhc_leung17}
{Leung}, G.~C.~K., {Coil}, A.~L., {Azadi}, M., {et~al.} 2017, \apj, 849, 48

\bibitem[{{Lilly} {et~al.}(2013){Lilly}, {Carollo}, {Pipino}, {Renzini}, \&
  {Peng}}]{rhc_lilly13}
{Lilly}, S.~J., {Carollo}, C.~M., {Pipino}, A., {Renzini}, A., \& {Peng}, Y.
  2013, \apj, 772, 119

\bibitem[{{Maiolino} {et~al.}(2012){Maiolino}, {Gallerani}, {Neri}, {Cicone},
  {Ferrara}, {Genzel}, {Lutz}, {Sturm}, {Tacconi}, {Walter}, {Feruglio},
  {Fiore}, \& {Piconcelli}}]{rhc_maiolino12}
{Maiolino}, R., {Gallerani}, S., {Neri}, R., {et~al.} 2012, \mnras, 425, L66

\bibitem[{{McMullin} {et~al.}(2007){McMullin}, {Waters}, {Schiebel}, {Young},
  \& {Golap}}]{rhc_casa}
{McMullin}, J.~P., {Waters}, B., {Schiebel}, D., {Young}, W., \& {Golap}, K.
  2007, in Astronomical Society of the Pacific Conference Series, Vol. 376,
  Astronomical Data Analysis Software and Systems XVI, ed. R.~A. {Shaw},
  F.~{Hill}, \& D.~J. {Bell}, 127

\bibitem[{{Morganti} {et~al.}(2016){Morganti}, {Veilleux}, {Oosterloo}, {Teng},
  \& {Rupke}}]{rhc_morganti16}
{Morganti}, R., {Veilleux}, S., {Oosterloo}, T., {Teng}, S.~H., \& {Rupke}, D.
  2016, \aap, 593, A30

\bibitem[{{Moster} {et~al.}(2013){Moster}, {Naab}, \& {White}}]{rhc_moster13}
{Moster}, B.~P., {Naab}, T., \& {White}, S.~D.~M. 2013, \mnras, 428, 3121

\bibitem[{{Muratov} {et~al.}(2015){Muratov}, {Kere{\v s}},
  {Faucher-Gigu{\`e}re}, {Hopkins}, {Quataert}, \& {Murray}}]{rhc_muratov15}
{Muratov}, A.~L., {Kere{\v s}}, D., {Faucher-Gigu{\`e}re}, C.-A., {et~al.}
  2015, \mnras, 454, 2691

\bibitem[{{Murphy} {et~al.}(2011){Murphy}, {Condon}, {Schinnerer}, {Kennicutt},
  {Calzetti}, {Armus}, {Helou}, {Turner}, {Aniano}, {Beir{\~a}o}, {Bolatto},
  {Brandl}, {Croxall}, {Dale}, {Donovan Meyer}, {Draine}, {Engelbracht},
  {Hunt}, {Hao}, {Koda}, {Roussel}, {Skibba}, \& {Smith}}]{rhc_murphy11}
{Murphy}, E.~J., {Condon}, J.~J., {Schinnerer}, E., {et~al.} 2011, \apj, 737,
  67

\bibitem[{{Muzzin} {et~al.}(2013){Muzzin}, {Marchesini}, {Stefanon}, {Franx},
  {McCracken}, {Milvang-Jensen}, {Dunlop}, {Fynbo}, {Brammer}, {Labb{\'e}}, \&
  {van Dokkum}}]{rhc_muzzin13}
{Muzzin}, A., {Marchesini}, D., {Stefanon}, M., {et~al.} 2013, \apj, 777, 18

\bibitem[{{Nardini} {et~al.}(2015){Nardini}, {Reeves}, {Gofford}, {Harrison},
  {Risaliti}, {Braito}, {Costa}, {Matzeu}, {Walton}, {Behar}, {Boggs},
  {Christensen}, {Craig}, {Hailey}, {Matt}, {Miller}, {O'Brien}, {Stern},
  {Turner}, \& {Ward}}]{rhc_nardini15}
{Nardini}, E., {Reeves}, J.~N., {Gofford}, J., {et~al.} 2015, Science, 347, 860

\bibitem[{{Nelson} {et~al.}(2016){Nelson}, {van Dokkum}, {F{\"o}rster
  Schreiber}, {Franx}, {Brammer}, {Momcheva}, {Wuyts}, {Whitaker}, {Skelton},
  {Fumagalli}, {Hayward}, {Kriek}, {Labb{\'e}}, {Leja}, {Rix}, {Tacconi}, {van
  der Wel}, {van den Bosch}, {Oesch}, {Dickey}, \& {Ulf Lange}}]{rhc_nelson16}
{Nelson}, E.~J., {van Dokkum}, P.~G., {F{\"o}rster Schreiber}, N.~M., {et~al.}
  2016, \apj, 828, 27

\bibitem[{{Newman} {et~al.}(2013){Newman}, {Genzel}, {F{\"o}rster Schreiber},
  {Shapiro Griffin}, {Mancini}, {Lilly}, {Renzini}, {Bouch{\'e}}, {Burkert},
  {Buschkamp}, {Carollo}, {Cresci}, {Davies}, {Eisenhauer}, {Genel}, {Hicks},
  {Kurk}, {Lutz}, {Naab}, {Peng}, {Sternberg}, {Tacconi}, {Wuyts}, {Zamorani},
  \& {Vergani}}]{rhc_newman13}
{Newman}, S.~F., {Genzel}, R., {F{\"o}rster Schreiber}, N.~M., {et~al.} 2013,
  \apj, 767, 104

\bibitem[{{Newman} {et~al.}(2014){Newman}, {Buschkamp}, {Genzel}, {F{\"o}rster
  Schreiber}, {Kurk}, {Sternberg}, {Gnat}, {Rosario}, {Mancini}, {Lilly},
  {Renzini}, {Burkert}, {Carollo}, {Cresci}, {Davies}, {Eisenhauer}, {Genel},
  {Shapiro Griffin}, {Hicks}, {Lutz}, {Naab}, {Peng}, {Tacconi}, {Wuyts},
  {Zamorani}, {Vergani}, \& {Weiner}}]{rhc_newman14}
{Newman}, S.~F., {Buschkamp}, P., {Genzel}, R., {et~al.} 2014, \apj, 781, 21

\bibitem[{{Oppenheimer} {et~al.}(2010){Oppenheimer}, {Dav{\'e}}, {Kere{\v s}},
  {Fardal}, {Katz}, {Kollmeier}, \& {Weinberg}}]{rhc_oppenheimer10}
{Oppenheimer}, B.~D., {Dav{\'e}}, R., {Kere{\v s}}, D., {et~al.} 2010, \mnras,
  406, 2325

\bibitem[{{Pereira-Santaella} {et~al.}(2018){Pereira-Santaella}, {Colina},
  {Garcia-Burillo}, {Combes}, {Emonts}, {Aalto}, {Alonso-Herrero}, {Arribas},
  {Henkel}, {Labiano}, {Muller}, {Piqueras Lopez}, {Rigopoulou}, \& {van der
  Werf}}]{rhc_p-s18}
{Pereira-Santaella}, M., {Colina}, L., {Garcia-Burillo}, S., {et~al.} 2018,
  ArXiv e-prints, arXiv:1805.03667

\bibitem[{{Pettini} \& {Pagel}(2004)}]{rhc_pp04}
{Pettini}, M., \& {Pagel}, B.~E.~J. 2004, \mnras, 348, L59

\bibitem[{{Richings} \& {Faucher-Gigu{\`e}re}(2018)}]{rhc_richings18}
{Richings}, A.~J., \& {Faucher-Gigu{\`e}re}, C.-A. 2018, \mnras, 474, 3673

\bibitem[{{Rodr{\'{\i}}guez Zaur{\'{\i}}n} {et~al.}(2013){Rodr{\'{\i}}guez
  Zaur{\'{\i}}n}, {Tadhunter}, {Rose}, \& {Holt}}]{rhc_r-z13}
{Rodr{\'{\i}}guez Zaur{\'{\i}}n}, J., {Tadhunter}, C.~N., {Rose}, M., \&
  {Holt}, J. 2013, \mnras, 432, 138

\bibitem[{{Rupke} {et~al.}(2002){Rupke}, {Veilleux}, \&
  {Sanders}}]{rhc_rupke02}
{Rupke}, D.~S., {Veilleux}, S., \& {Sanders}, D.~B. 2002, \apj, 570, 588

\bibitem[{{Rupke} {et~al.}(2005){Rupke}, {Veilleux}, \&
  {Sanders}}]{rhc_rupke05}
---. 2005, \apj, 632, 751

\bibitem[{{Rupke} \& {Veilleux}(2011)}]{rhc_rupke11}
{Rupke}, D.~S.~N., \& {Veilleux}, S. 2011, \apjl, 729, L27

\bibitem[{{Schawinski} {et~al.}(2015){Schawinski}, {Koss}, {Berney}, \&
  {Sartori}}]{rhc_schawinski15}
{Schawinski}, K., {Koss}, M., {Berney}, S., \& {Sartori}, L.~F. 2015, \mnras,
  451, 2517

\bibitem[{{Schinnerer} {et~al.}(2010){Schinnerer}, {Sargent}, {Bondi}, {Smol{\v
  c}i{\'c}}, {Datta}, {Carilli}, {Bertoldi}, {Blain}, {Ciliegi}, {Koekemoer},
  \& {Scoville}}]{rhc_schinnerer10}
{Schinnerer}, E., {Sargent}, M.~T., {Bondi}, M., {et~al.} 2010, \apjs, 188, 384

\bibitem[{{Silk} \& {Rees}(1998)}]{rhc_silk98}
{Silk}, J., \& {Rees}, M.~J. 1998, \aap, 331, L1

\bibitem[{{Skelton} {et~al.}(2014){Skelton}, {Whitaker}, {Momcheva}, {Brammer},
  {van Dokkum}, {Labb{\'e}}, {Franx}, {van der Wel}, {Bezanson}, {Da Cunha},
  {Fumagalli}, {F{\"o}rster Schreiber}, {Kriek}, {Leja}, {Lundgren}, {Magee},
  {Marchesini}, {Maseda}, {Nelson}, {Oesch}, {Pacifici}, {Patel}, {Price},
  {Rix}, {Tal}, {Wake}, \& {Wuyts}}]{rhc_skelton14}
{Skelton}, R.~E., {Whitaker}, K.~E., {Momcheva}, I.~G., {et~al.} 2014, \apjs,
  214, 24

\bibitem[{{Springel} {et~al.}(2005){Springel}, {Di Matteo}, \&
  {Hernquist}}]{rhc_springel05}
{Springel}, V., {Di Matteo}, T., \& {Hernquist}, L. 2005, \apjl, 620, L79

\bibitem[{{Sturm} {et~al.}(2011){Sturm}, {Gonz{\'a}lez-Alfonso}, {Veilleux},
  {Fischer}, {Graci{\'a}-Carpio}, {Hailey-Dunsheath}, {Contursi}, {Poglitsch},
  {Sternberg}, {Davies}, {Genzel}, {Lutz}, {Tacconi}, {Verma}, {Maiolino}, \&
  {de Jong}}]{rhc_sturm11}
{Sturm}, E., {Gonz{\'a}lez-Alfonso}, E., {Veilleux}, S., {et~al.} 2011, \apjl,
  733, L16

\bibitem[{{Tacchella} {et~al.}(2016){Tacchella}, {Dekel}, {Carollo},
  {Ceverino}, {DeGraf}, {Lapiner}, {Mandelker}, \& {Primack
  Joel}}]{rhc_tacchella16}
{Tacchella}, S., {Dekel}, A., {Carollo}, C.~M., {et~al.} 2016, \mnras, 457,
  2790

\bibitem[{{Tacchella} {et~al.}(2015{\natexlab{a}}){Tacchella}, {Carollo},
  {Renzini}, {Schreiber}, {Lang}, {Wuyts}, {Cresci}, {Dekel}, {Genzel},
  {Lilly}, {Mancini}, {Newman}, {Onodera}, {Shapley}, {Tacconi}, {Woo}, \&
  {Zamorani}}]{rhc_tacchella15b}
{Tacchella}, S., {Carollo}, C.~M., {Renzini}, A., {et~al.} 2015{\natexlab{a}},
  Science, 348, 314

\bibitem[{{Tacchella} {et~al.}(2015{\natexlab{b}}){Tacchella}, {Lang},
  {Carollo}, {F{\"o}rster Schreiber}, {Renzini}, {Shapley}, {Wuyts}, {Cresci},
  {Genzel}, {Lilly}, {Mancini}, {Newman}, {Tacconi}, {Zamorani}, {Davies},
  {Kurk}, \& {Pozzetti}}]{rhc_tacchella15}
{Tacchella}, S., {Lang}, P., {Carollo}, C.~M., {et~al.} 2015{\natexlab{b}},
  \apj, 802, 101

\bibitem[{{Tacconi} {et~al.}(2010){Tacconi}, {Genzel}, {Neri}, {Cox}, {Cooper},
  {Shapiro}, {Bolatto}, {Bouch{\'e}}, {Bournaud}, {Burkert}, {Combes},
  {Comerford}, {Davis}, {Schreiber}, {Garcia-Burillo}, {Gracia-Carpio}, {Lutz},
  {Naab}, {Omont}, {Shapley}, {Sternberg}, \& {Weiner}}]{rhc_tacconi10}
{Tacconi}, L.~J., {Genzel}, R., {Neri}, R., {et~al.} 2010, \nat, 463, 781

\bibitem[{{Tacconi} {et~al.}(2013){Tacconi}, {Neri}, {Genzel}, {Combes},
  {Bolatto}, {Cooper}, {Wuyts}, {Bournaud}, {Burkert}, {Comerford}, {Cox},
  {Davis}, {F{\"o}rster Schreiber}, {Garc{\'{\i}}a-Burillo}, {Gracia-Carpio},
  {Lutz}, {Naab}, {Newman}, {Omont}, {Saintonge}, {Shapiro Griffin}, {Shapley},
  {Sternberg}, \& {Weiner}}]{rhc_tacconi13}
{Tacconi}, L.~J., {Neri}, R., {Genzel}, R., {et~al.} 2013, \apj, 768, 74

\bibitem[{{Tacconi} {et~al.}(2018){Tacconi}, {Genzel}, {Saintonge}, {Combes},
  {Garc{\'{\i}}a-Burillo}, {Neri}, {Bolatto}, {Contini}, {F{\"o}rster
  Schreiber}, {Lilly}, {Lutz}, {Wuyts}, {Accurso}, {Boissier}, {Boone},
  {Bouch{\'e}}, {Bournaud}, {Burkert}, {Carollo}, {Cooper}, {Cox}, {Feruglio},
  {Freundlich}, {Herrera-Camus}, {Juneau}, {Lippa}, {Naab}, {Renzini},
  {Salome}, {Sternberg}, {Tadaki}, {{\"U}bler}, {Walter}, {Weiner}, \&
  {Weiss}}]{rhc_tacconi18}
{Tacconi}, L.~J., {Genzel}, R., {Saintonge}, A., {et~al.} 2018, \apj, 853, 179

\bibitem[{{Tecza} {et~al.}(2001){Tecza}, {Thatte}, \& {Maiolino}}]{rhc_tecza01}
{Tecza}, M., {Thatte}, N., \& {Maiolino}, R. 2001, in IAU Symposium, Vol. 205,
  Galaxies and their Constituents at the Highest Angular Resolutions, ed. R.~T.
  {Schilizzi}, 216

\bibitem[{{Tombesi} {et~al.}(2015){Tombesi}, {Mel{\'e}ndez}, {Veilleux},
  {Reeves}, {Gonz{\'a}lez-Alfonso}, \& {Reynolds}}]{rhc_tombesi15}
{Tombesi}, F., {Mel{\'e}ndez}, M., {Veilleux}, S., {et~al.} 2015, \nat, 519,
  436

\bibitem[{{{\"U}bler} {et~al.}(2014){{\"U}bler}, {Naab}, {Oser}, {Aumer},
  {Sales}, \& {White}}]{rhc_uebler14}
{{\"U}bler}, H., {Naab}, T., {Oser}, L., {et~al.} 2014, \mnras, 443, 2092

\bibitem[{{van de Voort} {et~al.}(2011){van de Voort}, {Schaye}, {Booth},
  {Haas}, \& {Dalla Vecchia}}]{rhc_vdv11}
{van de Voort}, F., {Schaye}, J., {Booth}, C.~M., {Haas}, M.~R., \& {Dalla
  Vecchia}, C. 2011, \mnras, 414, 2458

\bibitem[{{Veilleux} {et~al.}(2017){Veilleux}, {Bolatto}, {Tombesi},
  {Mel{\'e}ndez}, {Sturm}, {Gonz{\'a}lez-Alfonso}, {Fischer}, \&
  {Rupke}}]{rhc_veilleux17}
{Veilleux}, S., {Bolatto}, A., {Tombesi}, F., {et~al.} 2017, \apj, 843, 18

\bibitem[{{Veilleux} {et~al.}(2005){Veilleux}, {Cecil}, \&
  {Bland-Hawthorn}}]{rhc_veilleux05}
{Veilleux}, S., {Cecil}, G., \& {Bland-Hawthorn}, J. 2005, \araa, 43, 769

\bibitem[{{Veilleux} {et~al.}(2009){Veilleux}, {Rupke}, {Kim}, {Genzel},
  {Sturm}, {Lutz}, {Contursi}, {Schweitzer}, {Tacconi}, {Netzer}, {Sternberg},
  {Mihos}, {Baker}, {Mazzarella}, {Lord}, {Sanders}, {Stockton}, {Joseph}, \&
  {Barnes}}]{rhc_veilleux09}
{Veilleux}, S., {Rupke}, D.~S.~N., {Kim}, D.-C., {et~al.} 2009, \apjs, 182, 628

\bibitem[{{Veilleux} {et~al.}(2013){Veilleux}, {Mel{\'e}ndez}, {Sturm},
  {Gracia-Carpio}, {Fischer}, {Gonz{\'a}lez-Alfonso}, {Contursi}, {Lutz},
  {Poglitsch}, {Davies}, {Genzel}, {Tacconi}, {de Jong}, {Sternberg}, {Netzer},
  {Hailey-Dunsheath}, {Verma}, {Rupke}, {Maiolino}, {Teng}, \&
  {Polisensky}}]{rhc_veilleux13}
{Veilleux}, S., {Mel{\'e}ndez}, M., {Sturm}, E., {et~al.} 2013, \apj, 776, 27

\bibitem[{{Walter} {et~al.}(2017){Walter}, {Bolatto}, {Leroy}, {Veilleux},
  {Warren}, {Hodge}, {Levy}, {Meier}, {Ostriker}, {Ott}, {Rosolowsky},
  {Scoville}, {Weiss}, {Zschaechner}, \& {Zwaan}}]{rhc_walter17}
{Walter}, F., {Bolatto}, A.~D., {Leroy}, A.~K., {et~al.} 2017, \apj, 835, 265

\bibitem[{{Woo} {et~al.}(2016){Woo}, {Bae}, {Son}, \& {Karouzos}}]{rhc_woo16}
{Woo}, J.-H., {Bae}, H.-J., {Son}, D., \& {Karouzos}, M. 2016, \apj, 817, 108

\bibitem[{{Zubovas} \& {King}(2012)}]{rhc_zubovas12}
{Zubovas}, K., \& {King}, A. 2012, \apjl, 745, L34

\bibitem[{{Zubovas} \& {King}(2016)}]{rhc_zubovas16}
---. 2016, \mnras, 462, 4055

\bibitem[{{Zubovas} \& {King}(2014)}]{rhc_zubovas14a}
{Zubovas}, K., \& {King}, A.~R. 2014, \mnras, 439, 400

\bibitem[{{Zubovas} \& {Nayakshin}(2014)}]{rhc_zubovas14b}
{Zubovas}, K., \& {Nayakshin}, S. 2014, \mnras, 440, 2625

\end{thebibliography}

\end{document}